\documentclass[
aps,
pra,
showpacs
amsmath,
amssymb,
preprint,
floatfix,
lengthcheck, 
superscriptaddress,
longbibliography
]{revtex4-1}

\usepackage{graphicx}
\usepackage{mathtools}
\usepackage{braket}
\usepackage{hyperref}
\usepackage{bbold}
\usepackage{calrsfs}
\usepackage{dsfont}
\usepackage{cancel}
\usepackage{mathrsfs}
\usepackage[normalem]{ulem}

\usepackage{color}

\begin{document}


\title{Non-perturbative Mass Renormalization Effects in \\ Non-relativistic Quantum Electrodynamics }

\author{Davis M. Welakuh}
\email[Electronic address:\;]{davis.welakuh@mpsd.mpg.de}
\affiliation{Simons Center for Computational Physical Chemistry at New York University, New York, New York 10003, USA}
\affiliation{Max Planck Institute for the Structure and Dynamics of Matter, Luruper Chaussee 149, 22761 Hamburg, Germany}

\author{Vasil Rokaj}
\email[Electronic address:\;]{vasil.rokaj@villanova.edu}
\affiliation{ITAMP, Center for Astrophysics $|$ Harvard \& Smithsonian, Cambridge, Massachusetts 02138, USA}
\affiliation{Department of Physics, Harvard University, Cambridge, Massachusetts 02138, USA}
\affiliation{Department of Physics, Villanova University, Villanova, Pennsylvania 19085, USA \looseness=-1}

\author{Michael Ruggenthaler}
\email[Electronic address:\;]{michael.ruggenthaler@mpsd.mpg.de}
\affiliation{Max Planck Institute for the Structure and Dynamics of Matter, Luruper Chaussee 149, 22761 Hamburg, Germany}
\affiliation{The Hamburg Center for Ultrafast Imaging, Luruper Chaussee 149, 22761 Hamburg, Germany}

\author{Angel Rubio}
\email[Electronic address:\;]{angel.rubio@mpsd.mpg.de}
\affiliation{Max Planck Institute for the Structure and Dynamics of Matter, Luruper Chaussee 149, 22761 Hamburg, Germany}
\affiliation{Center for Computational Quantum Physics, Flatiron Institute, 162 5th Avenue, New York, NY 10010, USA}


\begin{abstract}
In this work we investigate the effects that multi-mode photonic environments, e.g., optical cavities, have on the properties of quantum matter. We highlight the importance of the non-perturbative mass renormalization procedure for ab initio quantum electrodynamics simulations and how it connects to common approximations used in polaritonic chemistry and cavity materials engineering. We focus on one-dimensional systems which can be solved exactly for large number of photon modes. First, we apply mass renormalization to free particles. The value of the renormalized mass depends on the details of the photonic environment and on the number of particles. We then show how the multi-mode photon field influences various ground- and excited-state properties of atomic and molecular systems. For instance, {we observe the enhancement of particle confinement in the binding potential for the atomic system}, and the modification of the potential energy surfaces of the molecular dimer due to photon-mediated long-range interactions. We also highlight how these changes compare to the common free-space mass-renormalization approximation employed in electronic structure theory and quantum chemistry. Since such phenomena are enhanced under strong light-matter coupling in a cavity environment they will become relevant for the emerging fields of polaritonic chemistry and cavity materials engineering.
\end{abstract}

\maketitle


\section{Introduction}
\label{sec:introduction}


In recent years, a multitude of seminal experimental and theoretical breakthroughs involving atoms, molecules, and solids embedded in photonic structures have ushered in the rapidly developing fields of polaritonic chemistry~\cite{ruggenthaler2017b,flick2018a,ruggenthaler2023} and cavity quantum materials~\cite{hubener2021,schlawin2022,ebbesen2016,garcia-vidal2021}. The most important and striking aspect of these fields is the capability of modifying or controlling the properties of matter in an unprecedented way by coupling it strongly to the vacuum modes of a photonic structure. Some examples of the experimental and
theoretical works include the possibility of building polariton lasers~\cite{cohen2010}, control photochemical reactions~\cite{hutchison2012,galego2016,sidler2020} and energy transfer~\cite{coles2014,schaefer2019,zhong2016}, enhancement of harmonic generation from polaritonic states~\cite{chervy2016,barachati2018,welakuh2022b,welakuh2022c}, modification of ground-state chemical reactions via vibrational strong coupling~\cite{thompson2006,schaefer2022,thomas2019}, or cavity-control {of condensed matter properties}~\cite{peter2005,latini2019,latini2021,liu2015, FaistCavityHall, paravacini2019, RokajButterfly2022, rokaj2023topological,vinas2023,sentef2018}. Also, the coupled light-matter system can be tuned to exhibit significantly different properties even at room temperature~\cite{chikkaraddy2016,ojambati2019}. The variety of these different effects (which is by no means a comprehensive list) shows the complexity that results from a strongly coupled light-matter system. It is clear that the theoretical description of these effects is far from trivial as it requires knowledge and methods ranging from materials science, quantum chemistry, quantum optics and many-body physics.

The theoretical tools employed to explain the experimental results {are typically} quantum optical models (e.g., Tavis-Cummings or Dicke-model)~\cite{dicke1954,tavis1968} or perturbative methods similar to scattering considerations in high-energy physics~\cite{greiner1996,ryder1996}, and only recently first-principles approaches for coupled light-matter systems have been developed~\cite{ruggenthaler2014,flick2015,haugland2020,riso2022}. Usually these approaches in practice employ the few- or even single-effective-mode approximation~\cite{galego2017,gonzalez2016}. Also for macroscopic molecular ensembles and extended systems it has been recently shown how such an effective few-mode approximation can be properly defined~\cite{svendsen2023theory}. While such approximations are often well-justified, there are many effects that need a multi-mode description~\cite{ruggenthaler2023,svendsen2024ab, svendsen2023theory}. Examples include radiative dissipation and finite-lifetime effects, such as the Purcell effect~\cite{purcell1995}, dispersive forces (Casimir and van der Walls forces)~\cite{casimir1948} or renormalization effects such as the Lamb-shift~\cite{bethe1957}. In the case of few-level quantum optical models, important multi-mode phenomena have been studied in detail~\cite{weisskopf1930,buzek1999,flick2017}. However, due to the reduction of the matter degrees of freedom, typically mass renormalization effects are not considered. For example, only recently the renormalization group flow of the Jaynes-Cummings model was studied~\cite{JCrenormalization}.

For first-principles methods, it has been pointed out that the mass renormalization of charged particles {due to the transverse photonic degrees of freedom} can have important implications for physical systems~\cite{spohn2004, rokaj2020, ruggenthaler2023}. Given the fact that the electromagnetic environment in a cavity is drastically different than free space, because the photon density of states, electromagnetic spectrum and light-matter coupling are modified, it becomes necessary to investigate how multi-mode renormalization effects in the cavity emerge under strong coupling. {This is specifically important in the context of simulating multi-mode situations in polaritonic chemistry and cavity materials engineering, where usually the free-space renormalized mass of the charged particles is employed~\cite{ruggenthaler2023}. Such a procedure runs the risk of erroneously double-counting the free-space continuum of modes~\cite{svendsen2023theory}. Yet,} having in mind the recent successes in experimentally modifying chemical reactions and material properties, {such} effects could be part of the solution to the conundrum of how photon-field fluctuations can influence atoms, molecules and solid-state systems even for ambient conditions. We further note that similar renormalization effects are well-known in, e.g., solid-state physics, where the longitudinal Coulomb interaction leads to the emergence of new quasi-particles with effective/renormalized masses~\cite{kittel2018introduction}. We highlight, however, that here we focus on the mass renormalization due to the transverse photonic degrees of freedom. In order to quantify the effects on equilibrium states, a non-perturbative ab initio investigation that solves the coupled light-matter problem exactly is needed. For our working definition of ab initio in the context of coupled light-matter problems we refer the reader to appendix A of Ref.~\cite{ruggenthaler2023}.

{At this point it becomes important to stress the difference between mass renormalization and quasi-particles due to the Coulomb interaction between many charged particles, and the mass renormalization of a single, free charged particle due to the self-consistent interaction with the transverse photon field.} Already for a classical charged particle coupled to the classical electromagnetic field, the back-action of the self-field, i.e., the electromagnetic field that a charged particle generates due to its own presence, leads to the need to renormalize the mass of the particle~\cite{spohn2004}. Without introducing a smallest length scale, the self-field would trap the particle and it would not move. That is, already for merely having a theoretical description of the energy-momentum (dispersion) relation of a free charged particle, we need to regularize the ultraviolet behavior and introduce a corresponding \textit{bare} mass. {By considering the bare mass to be a function of the ultraviolet regularization} one can describe the observed dispersion relation. If we discard the influence of the self-field completely, i.e., we choose an ultraviolet cutoff zero, we call the mass in the energy-momentum relation the \textit{observable} mass. This problem survives the quantization of the matter and photon degrees of freedom~\cite{spohn2004, ryder1996, greiner1996} {and we note again that the ubiquitous mass renormalization and emergence of quasi-particles due to the longitudinal Coulomb interaction is a different effect. Moreover, in the context of multi-mode photonic environments in polaritonic chemistry and cavity materials engineering, where the ultraviolet cutoff is by construction non-zero, one then has to either work with the corresponding bare masses or one has to keep track of the difference to the free-space continuum~\cite{svendsen2023theory}. This subtlety is, however, usually ignored and the} question of how the bare mass ``runs" with the cutoff in the dispersion relation {moreover} depends on the theory {used to describe the physical system and is thus far from trivial}. For instance, in relativistic quantum electrodynamics (QED) the bare mass goes to infinity as the cutoff is increased {and we note that its ``running" can also be influenced by the respective anti-particles~\cite{karbstein2008}}. For non-relativistic QED (NRQED), where the electrons are assumed to have small momenta such that a Schr\"odinger description becomes appropriate, the bare mass goes to zero~\cite{spohn2004}. This is quite intuitive, since the bare mass appears in the denominator of the kinetic energy and in order to cancel the arbitrarily large self-energy the $1/m$ prefactor needs to diverge as well.

We note that to understand this ``running" of the mass {due to coupling to the transverse photon degrees of freedom} of a single free particle within a QED theory, we have two equivalent options: We can either fix the dispersion relation and consider how the bare mass needs to be adapted when increasing the cutoff, or we fix the bare mass and consider how the dispersion changes. Both considerations give us the same information, yet in the latter option we consider a hypothetical observable mass that changes with the cutoff. In this case the hypothetical observable mass is merely a theoretical tool.

It is obvious that this mass-renormalization procedure merely fixes the single-particle free-space dispersion relation, while it does not fix how other observables or systems are affected by changing the cutoffs within a QED theory. There is apriori no reason to believe that this dispersion-based mass-renormalization procedure gives the same results as when we would try to make, e.g., the one-particle density of a bound-state system of the bare and renormalized system the same. It is even unclear whether the dispersion-based mass renormalization of a single free particle gives the same relation as the dispersion-based mass renormalization for a single free particle in an ensemble of many free particles. {Moreover, it remains completely unexplored how this mass renormalization due to the transverse photon fields affects the emergence of quasi particles that arise due to the longitudinal Coulomb interaction.} In this work we want to explore such mass renormalization {due to the transverse photonic degrees of freedom and its interplay with the longitudinal Coulomb interaction} in the context of ab initio NRQED. The NRQED setting describes the quantum mechanical interaction between the constituents of matter (electrons and nuclei) and photon degrees alike.

Such a study is, however, far from trivial and we need to make some initial assumptions to make it tractable. Firstly, since we are mainly interested in the effect of optical wavelengths on bound-state systems, we make the long-wavelength approximation (also called dipole-approximation or optical limit~\cite{tannoudji1989,spohn2004}). This approximation is standard in most cases of polaritonic chemistry and material sciences~\cite{ruggenthaler2017b,schlawin2022,ebbesen2016,welakuh2021}. In the case of the long-wavelength approximation the relation between the observable and the bare mass in QED~\cite{spohn2004} is known analytically and non-perturbatively for a single free particle~\cite{hainzl2002, rokaj2020}\footnote{We note that in the work of \citet{hainzl2002} the minimal-coupling considerations are perturbative and are then contrasted to the non-perturbative (to all orders) results in the long-wavelength approximation known from the thesis of van Kampen.} This result allows us to verify the accuracy of our numerical simulations when going from free particles to bound-state systems. Secondly, in order to represent the continuum of modes numerically, we require a dense sampling of the relevant energy range. For a non-perturbative simulation this becomes exceptionally demanding for three-dimensions. We will therefore restrict to a one-dimensional continuum of modes and respective atomic and molecular models~\cite{flick2017,landau1977,albareda2021}. We will comment on the range of validity of these two assumptions and on the implications of our results for general situations later.

Having set the stage, {the aim of this work is to investigate non-perturbative mass renormalization due to the transverse photonic degrees of freedom in the emerging fields of polaritonic chemistry and cavity quantum materials. We demonstrate the mass renormalization first for a single particle in} free-space in non-perturbative QED. This introduces the concept of a \textit{bare mass} in a way that is most closely connected to the common renormalization procedure of perturbative QED~\cite{hainzl2002,mandl2010}. We demonstrate how the mass-renormalization procedure introduces the observable \textit{renormalized mass} and how this {connects} to the usual energy dispersion of {a single free particle in} quantum mechanics. In this simplest of situations we then highlight the need to go beyond perturbation theory {when we consider the coupled wave function and we uncover that the energy- and length-scales of the light and matter modes/states need to match to recover the analytic results in a non-perturbative simulation}. Next we show how the continuum of modes influences the ground-state of atomic and molecular systems. Interestingly we find that the renormalized (observable) mass approximation, as employed {in electronic structure theory and quantum chemistry}, shows relatively strong deviations from the full multi-mode simulations. The discrepancies become more pronounced when going to the molecular case. These results highlight an important feature of multi-mode light-matter interaction for bound matter systems, where the lower-lying modes of the sampled electromagnetic continuum couple more strongly than the higher-lying modes. We then show how modifying the electromagnetic vacuum by, e.g., an optical cavity, can affect the ground-state properties of atomic and molecular systems. For the atomic model we find that the ground-state density of the electron gets {more localized} due to the interaction with the multi-mode cavity field. For the molecular model we observe the modification of the ground-state potential energy surface (PES) due to cavity-mediated long-range interactions. We finally {connect our results to multi-mode simulations in polaritonic chemistry and cavity materials engineering and comment on} the general case of three-dimensional light-matter systems.

\section{Theoretical framework}
\label{sec:theory}

We will focus in the following on the non-relativistic limit of QED for the charged particles and assume that the wavelengths of the transverse modes of relevance are much larger than the exponentially localized matter system, such that the long-wavelength limit~\cite{tannoudji1989} is applicable. In this setting, the dynamics of the coupled system is described by the velocity (momentum) form of the Pauli-Fierz Hamiltonian~\cite{rokaj2017,spohn2004} 
\begin{align}
\hat{H}_{\text{V}} &= \sum\limits_{l=1}^{N_{e}}\frac{1}{2m} \left(- i\hbar\boldsymbol{\nabla}_{\textbf{r}_{l}} - \frac{|e|}{c} \hat{\textbf{A}}\right)^{2} + \frac{1}{2} \sum\limits_{l\neq j}^{N_{e}} w(|\hat{\textbf{r}}_{l} - \hat{\textbf{r}}_{j}|) \nonumber \\
& \quad + \sum\limits_{l=1}^{N_{n}}\frac{1}{2M_{l}} \left(- i\hbar\boldsymbol{\nabla}_{\textbf{R}_{l}} + \frac{Z_{l}|e|}{c}\hat{\textbf{A}}\right)^{2}  \nonumber \\
& \quad + \frac{1}{2} \sum\limits_{l\neq j}^{N_{n}} Z_{l}Z_{j}w(|\hat{\textbf{R}}_{l} - \hat{\textbf{R}}_{j}|) + \sum_{\alpha=1}^{N_{p}} \hbar \omega_{\alpha} \left(\hat{a}^{\dagger}_{\alpha} \hat{a}_{\alpha}+\tfrac{1}{2}\right) \nonumber \\
& \quad - \sum\limits_{l=1}^{N_{e}} \sum\limits_{j=1}^{N_{n}} Z_{j}w(|\hat{\textbf{r}}_{l} - \hat{\textbf{R}}_{j}|)  , \label{eq:velocity-gauge-hamiltonian}
\end{align}
where the positive parameters $m$ and $M_l$ are the \textit{bare masses} of the $N_{e}$ electrons and $N_{n}$ nuclei, respectively. These are \textit{not} the usual masses of quantum mechanics (compare with Eq.~\eqref{eq:matter-hamiltonian}), and the elucidation of the effect of these bare masses and their relation with the cutoffs in the context of NRQED is the central topic of this manuscript. The electrons and nuclei are respectively described by the coordinates, $\hat{\textbf{r}}_{l}$ and $\hat{\textbf{R}}_{l}$, and $w$ is the longitudinal interaction between the charged particles. In free space and in three dimensions it is the usual Coulomb interaction $w(|\hat{\textbf{r}} - \hat{\textbf{r}}'|)=e^{2}/4\pi\varepsilon_{0}|\hat{\textbf{r}} - \hat{\textbf{r}}'|$. The energy of the quantized electromagnetic field is given in terms of the photon creation $\hat{a}^{\dagger}_{\alpha}$ and annihilation $\hat{a}_{\alpha}$ operators with associated mode frequency $\omega_{\alpha}$ for each mode $\alpha$ of an arbitrarily large but finite number of photon modes $N_{p}$. The vector potential is 
\begin{equation}
 \hat{\textbf{A}}=\sum_{\alpha=1}^{N_{p}}c \,\boldsymbol{\lambda}_{\alpha} \sqrt{\frac{\hbar}{2\omega_{\alpha}}}\left(\hat{a}_{\alpha} + \hat{a}_{\alpha}^{\dagger}\right) \;\; \textrm{with}\;\;  \boldsymbol{\lambda}_{\alpha}=\sqrt{\frac{1}{\epsilon_{0}V_{\alpha}}}\textbf{e}_{\alpha}.
\end{equation}
{Note that} $\boldsymbol{\lambda}_{\alpha}$ is the vectorial coupling parameter and $V_{\alpha}$ is the mode volume of the mode $\alpha$. {We highlight that for} general photonic structures the proper definition of the mode volume is non-trivial and might depend even on the matter system under consideration~\cite{svendsen2023theory}. In the case of three-dimensional free space the coupling becomes proportional to the fine-structure constant~\cite{rokaj2020}. In the simple Fabry-P\'erot cavities the coupling becomes proportional to mirror distances and the finesse of the cavity~\cite{svendsen2023theory}. The collective index $\alpha \equiv (\textbf{k}s)$ is used to denote the photon wave vector and the two transverse polarization directions $s = 1, 2$. We further note that we follow the usual convention of constructive quantum field theories to consider a discretized continuum that converges in resolvent-norm to the full continuum solution~\cite{glimm1970lambda, arai1997existence,miyao2020note}. Moreover, we stress that when sampling the photon modes, we cannot go to arbitrary high photon momenta $|\textbf{k}|\rightarrow \infty$, otherwise the Pauli-Fierz Hamiltonian will be ill-defined~\cite{spohn2004}. This mathematical fact is easy to understand on physical grounds, specifically for the long-wavelength approximation, since arbitrarily high momenta directly contradict our initial assumption to not resolve arbitrarily small length scales. Any microscopic length scale would be resolved with arbitrarily high frequencies. To circumvent this, the contributions of the photon continuum needs to be regularized by introducing an ultraviolet regularization~\cite{spohn2004,hainzl2002}. The effect of this regularization on physical properties is a further central topic of this work and will be discussed in the following.

It is important to note that the coupled light-matter system can be studied using the unitarily equivalent form of Eq.~(\ref{eq:velocity-gauge-hamiltonian})~\cite{rokaj2017}, the length form of the Pauli-Fierz Hamiltonian given as
\begin{align} 
\hat{H}_{\text{L}} &=- \frac{\hbar^{2}}{2m} \sum\limits_{l=1}^{N_{e}}\boldsymbol{\nabla}_{\textbf{r}_{l}}^{2} - \sum\limits_{l=1}^{N_{n}}\frac{\hbar^{2}}{2M_{l}} \boldsymbol{\nabla}_{\textbf{R}_{l}}^{2} + \frac{1}{2} \sum\limits_{l\neq j}^{N_{e}}w(|\hat{\textbf{r}}_{l} - \hat{\textbf{r}}_{j}|)  \nonumber \\
& \quad + \frac{1}{2} \sum\limits_{l\neq j}^{N_{n}}Z_{l}Z_{j}w(|\hat{\textbf{R}}_{l} - \hat{\textbf{R}}_{j}|)  \!-\! \sum\limits_{l=1}^{N_{e}} \sum\limits_{j=1}^{N_{n}} Z_{j}w(|\hat{\textbf{r}}_{l} - \hat{\textbf{R}}_{j}|) \nonumber \\
& \quad + \frac{1}{2} \sum_{\alpha=1}^{M} \left[\hat{p}^2_{\alpha} + \omega^2_{\alpha}\left(\hat{q}_{\alpha} \!-\! \frac{\boldsymbol{\lambda}_{\alpha}}{\omega_{\alpha}} \cdot \hat{\boldsymbol{\mu}}  \right)^2\right] , \label{eq:length-gauge-hamiltonian}
\end{align}
where the total dipole is $\hat{\boldsymbol{\mu}} =-\sum_{l=1}^{N_{e}} |e| \, \hat{\textbf{r}}_{l} +  \sum_{l=1}^{N_{n}}Z_{l}e\hat{\textbf{R}}_{l}$, $\hat{q}_{\alpha}$ is the displacement coordinate and $\hat{p}_{\alpha} = -i\hbar \tfrac{\partial}{\partial \hat{q}_{\alpha}}$ its conjugate momentum. We can define new creation and annihilation operators also for the length gauge, but we note that they are \textit{not} the original photonic operators as defined above in the velocity gauge but are mixed light-matter objects~\cite{rokaj2017,schaefer2020}.

The explicit interaction of the matter degrees of freedom with the photons as in Eqs.~(\ref{eq:velocity-gauge-hamiltonian}) and (\ref{eq:length-gauge-hamiltonian}), requires that we work with the \textit{bare masses, $m$ and $M_l$, for the electrons and nuclei respectively, as it is usually done in QED~\cite{spohn2004, bethe1947}}. {In electronic structure theory and quantum chemistry, however,} the observable masses of the particles are used {and the transverse photon modes are discarded}. The Hamiltonian describing this setting of interacting electrons and nuclei is
\begin{align} 
\hat{H} &=- \frac{\hbar^{2}}{2m_{e}}\sum\limits_{l=1}^{N_{e}}\boldsymbol{\nabla}_{\textbf{r}_{l}}^{2} - \sum\limits_{l=1}^{N_{n}}\frac{\hbar^{2}}{2M_{n,l}} \boldsymbol{\nabla}_{\textbf{R}_{l}}^{2}  \nonumber \\
& \quad + \frac{1}{2} \sum\limits_{l\neq j}^{N_{e}} w(|\hat{\textbf{r}}_{l} - \hat{\textbf{r}}_{j}|) + \frac{1}{2} \sum\limits_{l\neq j}^{N_{n}}Z_{l}Z_{j}w(|\hat{\textbf{R}}_{l} - \hat{\textbf{R}}_{j}|) \nonumber \\
& \quad - \sum\limits_{l=1}^{N_{e}} \sum\limits_{j=1}^{N_{n}} Z_{j}w(|\hat{\textbf{r}}_{l} - \hat{\textbf{R}}_{j}|) \, , \label{eq:matter-hamiltonian}
\end{align}
where $m_{e}$ and $M_{n,l}$ are the \textit{renormalized} or \textit{observable masses} of the electrons and nuclei in free space, respectively. In standard formulations of QED the following relation is assumed between the bare and the observable masses~\cite{spohn2004,hainzl2002,craig1998}
\begin{align} 
m_{e} &= m + m_{pt} \, , \label{eq:physical-mass} \\
M_{n,l} &= M_l + M_{pt,l} \, . \label{eq:physical-mass-nu}
\end{align}
The photon-induced masses $m_{pt}$ and $M_{pt,l}$ are interpreted respectively as the masses acquired by the electrons and nuclei due to the interaction with the photon field~\cite{craig1998}. The bare masses of NRQED and the renormalized masses are related via the free-space energy-momentum relation (see Sec.~\ref{sec:free-electron} for details). How well these two descriptions agree for other observables and properties of different systems is the main topic of this work. At this point it is important to mention that the Hamiltonians defined in this section refer to general many-body systems in three dimensions. In what follows in order to have exactly solvable models we will, however, focus on one-dimensional atomic and molecular models which can be simulated exactly when coupled to the electromagnetic continuum of modes. In the atomic case (Sec.~\ref{subsec:atomic-system}) we will consider only the single-electron case, while for the molecule (Sec.~\ref{subsec:molecular-system}) we have two electrons and two positively charged nuclei. But before we do so, let us consider how typically the bare and the observable masses are related~\cite{craig1998,spohn2004,hainzl2002}.

\section{Free particles coupled to the electromagnetic continuum}
\label{sec:free-electron}

To elucidate how {the bare masses of} NRQED and {the renormalized/observable masses} are related we consider the dispersion relation of free charged particles. We will use the case of free electrons in the following, but note that we can merely replace the charges and bare masses in the different formulas and also find the corresponding forms for the nuclei. We describe the free electrons coupled to the photon modes using Eq.~\eqref{eq:velocity-gauge-hamiltonian}. Further, we neglect the Coulomb interaction between the electrons, $w(|\hat{\textbf{r}} - \hat{\textbf{r}}'|)=0$. We note that in this section and the atomic model (Sec.~\ref{subsec:atomic-system}) we will focus on the single-electron case, where this is automatically fulfilled. Given this assumption the electronic eigenstates are plane waves of the form $e^{\textrm{i}kx}$, and the non-perturbative spectrum for $N_{e}$ electrons coupled to the vacuum photons in the long-wavelength approximation can be obtained analytically~\cite{rokaj2020}. It takes the form
\begin{align}
E_{\textbf{k}}(N_{p}) &= \frac{\hbar^{2}}{2m}\left(\sum_{j=1}^{N_{e}}\textbf{k}_{j}^{2} - \frac{1}{N_{e}} \sum_{\alpha=1}^{N_{p}} \frac{\omega_{d}^{2}}{\Omega_{\alpha}^{2}} (\tilde{\textbf{e}}_{\alpha}\cdot\textbf{K})^{2} \right) \nonumber \\
& \quad + \sum_{\alpha=1}^{N_{p}}\hbar\Omega_{\alpha}\left(n_{\alpha} + \frac{1}{2}\right) \, ,  \label{eq:free-electron-photon-spectrum}
\end{align}
where $\Omega_{\alpha}$ and $\tilde{\textbf{e}}_{\alpha}$ are the new normal modes and the new polarization vectors, the diamagnetic frequency of the system is defined as {\begin{equation}
    \omega_{d} = \sqrt{\frac{N_{e}e^{2}}{\epsilon_{0}mV_{\alpha}}}=|\boldsymbol{\lambda}_{\alpha}|\sqrt{\frac{N_{e}e^{2}}{m}}
\end{equation}} and $\textbf{K}=\sum_{j=1}^{N_{e}}\textbf{k}_{j}$ is the sum of all electronic momenta. To obtain the {renormalized} dispersion relation for the non-interacting free electron gas from Eq.~\eqref{eq:free-electron-photon-spectrum}, we can subsume the contributions of the photonic degrees and its interaction with the electronic system into the observable mass and find
\begin{align}
E_{\textbf{k}} = \frac{\hbar^{2}}{2m_{e}}\sum_{j=1}^{N_{e}}\textbf{k}_{j}^{2}  \, .  \label{eq:free-electron-spectrum}
\end{align}
In NRQED, the renormalized mass for free electrons is defined via the energy dispersion of the electrons at the lowest relevant frequencies and is {formally by}~\cite{chen2008,frohlich2010}:
\begin{align}
m_{e} =  \left[\frac{1}{\hbar^{2}} \frac{\partial^{2} E_{\textbf{k}}(N_{p})}{\partial \textbf{k}_{i}^{2}} \right] ^{-1}  \,  \label{eq:renormalized-mass}
\end{align}
{evaluated at the scale of interest.} Next, applying Eq.~\eqref{eq:renormalized-mass} to Eq.~\eqref{eq:free-electron-photon-spectrum}, we obtain the renormalized mass for the free electron gas given by
\begin{align}
m_{e} &= \frac{m}{1 - g(N_{p})} \quad \text{where} \quad g(N_{p})=\frac{1}{N_{e}}\sum_{\alpha=1}^{N_{p}}\frac{\omega_{d}^{2}}{\Omega_{\alpha}^{2}} (\tilde{\textbf{e}}_{\alpha}\cdot\textbf{e}_{i})^{2}.\label{eq:renormalized-mass-derived}
\end{align}
We note that $i=x,y,z$ and $g(N_{p})$ is the total multi-mode coupling to the electromagnetic field. Equation (\ref{eq:renormalized-mass-derived}) provides an analytic expression  of the connection between the bare mass $m$ and observable mass $m_{e}$ from a non-perturbative description. As already highlighted in the introduction, we here see explicitly that the bare mass in NRQED goes to zero when the ultraviolet regularization is removed. For a locally isotropic and homogeneous density of modes, such as in three-dimensional free space, where $\omega_{n} = c|\textbf{n}|(2\pi/L)$ for $\textbf{n} \in \mathbb{Z}_{0}^{3}$ and $V_{\alpha} = V = L^3$ the full quantization volume, we can connect to well-known results from renormalization theory~\cite{spohn2004}. Indeed, for three-dimensional free space we recover the fine-structure dependent mass renormalization of long-wavelength-approximated NRQED~\cite{rokaj2020}. The cutoff (or some other form of regularization) is needed to avoid the divergence of the observable mass, which for a single electron in three dimensions is found to be at exceedingly high energies corresponding to the energy regime of quantum chromodynamics (QCD)~\cite{rokaj2020}. We will discuss the choice of cutoff and its implications below. It is important to stress that the multi-mode coupling $g(N_{p})$ to the photon modes approaches unity, $ g(N_{p}) \to 1$, resulting to a diverging $m_{e}$ in the cases that we consider~\cite{rokaj2020}. To tame the diverging $m_e$ in renormalization theory, the bare mass $m$ becomes cutoff-dependent and is promoted into $m(N_{p})$ such that to exactly cancel the diverging term $1/(1-g(N_{p}))$. For that purpose one takes $m(N_{p})=m_e (1-g(N_{p}))$ where $m_e$ is the observable electron mass. In addition we would like to highlight that, strictly speaking, in a general, non-isotropic photonic environment the observable mass would become direction dependent, as can be seen from Eq.~\eqref{eq:renormalized-mass}. We will, however, in the following consider one-dimensional models and hence will ignore this subtle yet important point, and only comment on it at the end of this work. In the following we will use adapted units (a.u.) such that $e=(4\pi\epsilon_{0})^{-1}=\hbar=m=1$. These units are \textit{not} atomic units since we choose the bare electronic mass $m$ to be equal to one, and hence the units are adapted to the cutoff/scale of the model. We therefore consider the situation, as discussed in Sec.~\ref{sec:introduction}, where we fix the bare mass and investigate a changing hypothetical observable mass. We are not interested in the actual value of the observable mass but rather in how the wave function and its observables ``run" with the cutoff.

\begin{figure}
\includegraphics[width=0.99\columnwidth]{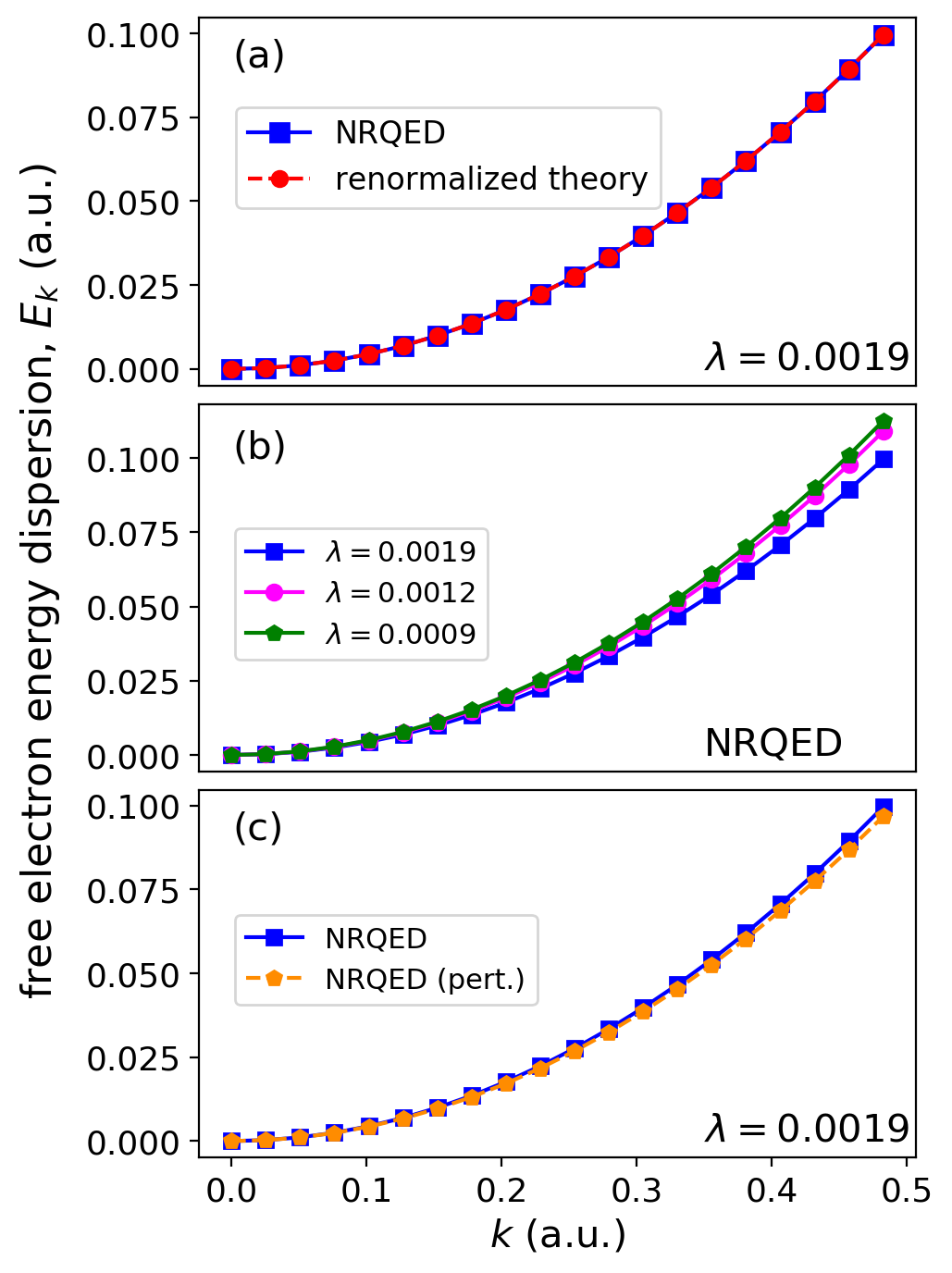}
\caption{(a) The energy dispersion of a free electron in the NRQED setting when coupled to 200 photon modes and the corresponding renormalized {description using the renormalized mass} $m_{e}$. Both settings show a quantitative agreement and we note this holds for different coupling strengths. (b) Comparison of the energy dispersion of NRQED for different $\lambda$'s where the dispersion relation becomes more flat increasing $\lambda$. (c) Comparison between the non-perturbative and perturbative (denoted pert.) NRQED energy dispersion, where the perturbative solution needs to employ also an infrared cutoff to lead to finite results. Still the perturbative results deviates for higher $k$ values.}
\label{fig:free-electron-M-modes}
\end{figure}

Now, we will consider a situation of a free electron restricted to one dimension interacting with a discretized electromagnetic continuum. With this model, we want to demonstrate the working principles of the mass-renormalization procedure and obtain the observable/renormalized mass of the interacting light-matter system. We choose the discretized photon continuum such that the range of its frequencies covers the desired energy range of the bound matter systems (discussed in Sec.~\ref{sec:bound-matter}). That is, the matter wave function of the bound-state system, which is equivalent to exponential localization~\cite{blanchard2003, spohn2004}, intrinsically sets the scale at which we investigate the present theory. In our case, we introduce lower and upper energy cutoffs which are respectively, 0.01 a.u. and 0.5 a.u. The upper cutoff is well within the validity of the dipole approximation. The lower cutoff is needed to treat the matter and the light sector consistently. Although non-perturbative NRQED has no infrared divergence~\cite{spohn2004}, the consistent treatment of the limit $\omega \rightarrow 0$ needs extra care and we discuss this in more detail at the end of this section. Here we choose the lower cutoff in agreement with the matter grid by having $l= 2\pi/k_{min}$. We sample the one-dimensional electromagnetic continuum by including explicitly 200 photon modes with equidistant energy spacing per mode of 0.00246 a.u. (see App.~\ref{sup:numerical-details} for details on the photon continuum). {This is smaller than the energy spacing of the matter system and choosing the sampling finer does not change the outcome, i.e., the results are converged.} This specified continuum of modes describes the local photonic density of states that we consider for the light-matter coupled system. In this setting of the coupled light-matter system, we compute the dispersion relations of Eq.~\eqref{eq:free-electron-photon-spectrum}, what we term as results from NRQED, and the {renormalized} dispersion relations of Eq.~\eqref{eq:free-electron-spectrum}.

\begin{figure}[bth]
\includegraphics[width=1.1\columnwidth]{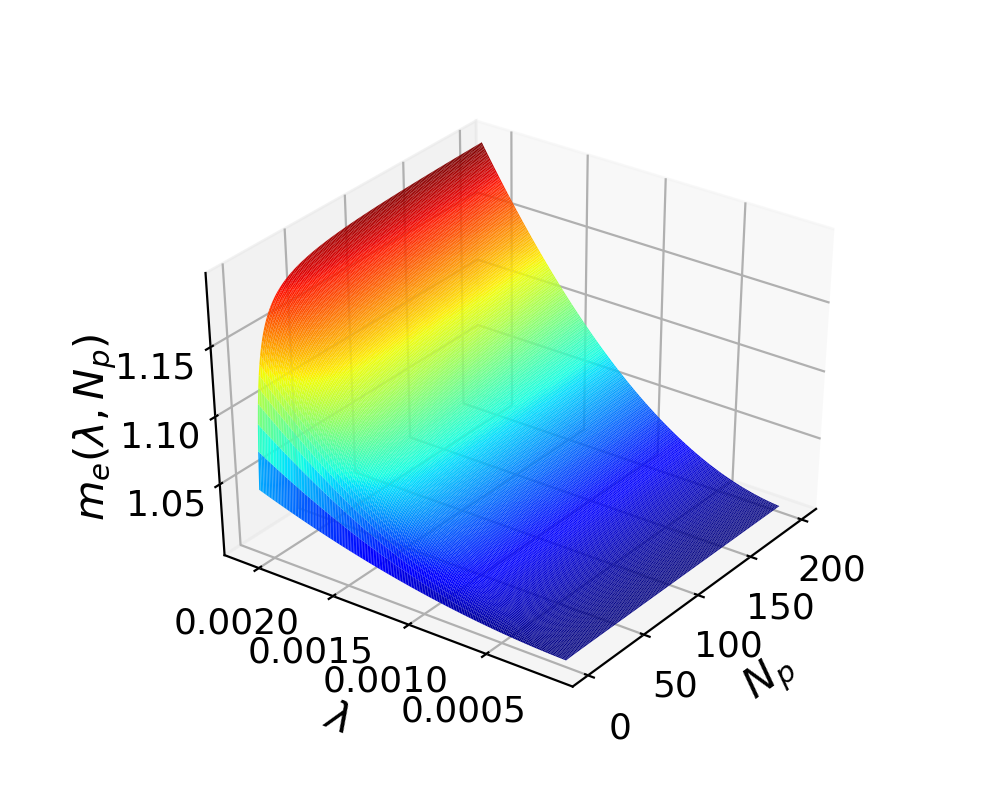}
\caption{The dependence of the renormalized mass on the coupling parameter $\lambda$ and number of photon modes $N_{p}$. For fixed coupling and increasing number of modes the mass increases quickly and asymptotically reaches a plateau at a finite value. For fixed number of modes and increasing coupling, the mass is shown to increase. }
\label{fig:mass-vs-coupling-modes}
\end{figure}

To obtain the {renormalized} dispersion relation requires that we perform a mass renormalization which accounts for the interaction between the bare free electron and the photon continuum. This procedure involves computing the observable mass as given in Eq.~\eqref{eq:renormalized-mass}. In Fig.~\ref{fig:free-electron-M-modes}a, we show a comparison of the energy dispersion obtained from {from both approaches}. We find that {(by construction) both approaches agree} which highlights a clear connection between those two settings when the photonic degrees are traced out and included in the renormalized mass. The results at the same time demonstrate the validity and working principles of the mass-renormalization procedure in the long-wavelength approximation. For the different coupling parameters $\lambda$, which for larger values indicate a stronger interaction with the photonic continuum, we find that the energy dispersion becomes more flat as shown in Fig.~\ref{fig:free-electron-M-modes}b. This is as a result of the photon-induced mass $m_{pt}$ as the free particle interacts with the photon field. Table~\eqref{tab:physical-masses} shows the hypothetical observable masses for different $\lambda$ and from which the photon-induced mass $m_{pt}$ can be deduced according to Eq.~\eqref{eq:physical-mass}. 
\begin{table}[bth]
\begin{tabular}{ | c | c | c |  }
	\hline
	  Coupling strength, $\lambda \;$  & $m_{e}$ &   $\;m_{e}$ (pert.)\\
	\hline 
	  $0.0019$       &  $1.1683661411$  & $1.2024522443$   \\
	\hline
	$0.0012$       &  $1.0673464565$  & $1.0722095112$  \\
	\hline
	$0.0009$       &  $1.0336732282$  & $1.0348466266$  \\
	\hline
\end{tabular}
\caption{The values of the renormalized mass obtained from Eq.~\eqref{eq:renormalized-mass-derived} for the non-perturbative and perturbative (denoted pert.) calculation for different coupling strengths $\lambda$ when coupled to 200 photon modes. The value of the bare mass used is $m=1$.}
\label{tab:physical-masses}
\end{table}

In Fig.~\ref{fig:mass-vs-coupling-modes} we show how the observable mass depends on the coupling parameter $\lambda$ and the number of photon modes $N_{p}$ (i.e. increasing photonic density of states and energy cutoff). We find that for a fixed light-matter coupling (e.g., $\lambda=0.0020$) and a non-zero lower photon cutoff, the renormalized mass increases as a function of $N_p$ and reaches a plateau without diverging. The renormalized mass diverges only in the case where the lower photon frequency is taken to zero. This point is discussed in more detail in App.~\ref{sup:nonperturbative-vs-perturbative} and is due to the one-dimensional setting considered in this work. In three-dimensions this effect is absent for isotropic and homogeneous modes due to the three-dimensional volume element~\cite{rokaj2020}.

At this point it becomes important to discuss the relation with a perturbative treatment. Up to second-order perturbation theory in terms of the electronic charge $e$ (see App.~\ref{sup:nonperturbative-vs-perturbative} for the continuum), the spectrum of an electron coupled to the discretized electromagnetic continuum is given by
\begin{align}
E_{k}^{(\text{pert})} &= \frac{k^{2}}{2}\left(1 - \sum_{n=1}^{\infty} \frac{\omega_{d}^{2}}{\omega_{n}^{2}} \right)  \, ,  \label{eq:free-electron-photon-spectrum-pert}
\end{align}
where the photon frequencies are $\omega_{n} =  2 \pi c n/L$. Using Euler's formula $\sum^{\infty}_{n=1}1/n^2=\pi^2/6$, and substituting the definition for the single-particle diamagnetic frequency $\omega_{d}^{2}=4 \pi /V$ and assuming for the effective quantization volume a cavity-like geometry of the form $V=AL$, where $A$ is the area of the mirrors and $L$ the mirror distance, we find the perturbative free particle dispersion
\begin{align}\label{eq:discretizedperturabtion}
E_{k}^{(\text{pert})} &= \frac{k^{2}}{2}\left(1 - \gamma \right) \;\; \textrm{where}\;\; \gamma=\frac{\pi L}{6 c^2 A} \, ,  
\end{align}
From the above result it becomes evident that if $\gamma>1$ then the free particle dispersion and the renormalized mass turn negative, signaling a break-down of perturbation theory. It is crucial to mention that the instability occurs for large $L$ which implies for low photon frequencies, i.e., perturbation theory becomes unstable in the infrared part of the spectrum. This is a striking result as it demonstrates that perturbation theory can become invalid even at low energies. In contrast, the non-perturbative multi-mode coupling from the exact solution has an upper bound and it does not exceed unity, $g(N_p)\leq 1$ (see App.~\ref{sup:nonperturbative-vs-perturbative}). Thus, the non-perturbative dispersion of the free particle is always positive and the system remains stable~\cite{rokaj2020}. This supports the idea by Van Hove~\cite{VANHOVE1952145} that divergences (or instabilities) in quantum field theories might not be a generic property but only due to perturbation theory. At this point it is important to mention that if we assume a cubic geometry for the quantization volume $V^{\prime}=L^3$, with $L$ the box-size, then the corresponding parameter which modifies the perturbative dispersion takes the form $\gamma^{\prime}=\frac{\pi }{6 c^2 L}$. In this case the instability point $\gamma^{\prime}>1$ of the perturbative dispersion occurs for small $L$, i.e., for high photonic frequencies. This is the standard ultraviolet diverging behavior of (three-dimensional) free space~\cite{craig1998, rokaj2020}. In order to make a comparison between non-perturbative NRQED and perturbative light-matter coupling possible, we need to avoid this infrared instability that only appears in perturbation theory. However, even if we do so, the corresponding lowest-order perturbative wave function~\footnote{We note that here we focus on just one out of the many modes that constitute the discretized continuum. We can re-construct the many-mode wave function by combining all possible excitation. Yet, if already the single-mode perturbative wave function can become inaccurate, the many-mode perturbative wave function will be so as well.}
\begin{align}
\ket{\Psi^{\rm (pert.)}} \approx \mathcal{C} \left(\ket{k} \otimes \ket{0} + \frac{\omega_d}{\omega} \frac{k}{\sqrt{2 \omega}} \ket{k}\otimes\ket{1}\right), 
\end{align}
where $\mathcal{C}$ is a normalization constant, is only accurate as long as the matter momentum scale is comparable with the photon momentum scale. That means, not only do we have a breakdown of perturbation theory if $\omega$ becomes too small for fixed $|k|$, but also if $|k|$ becomes too large for a fixed $\omega$. In addition we observe important differences between the perturbative and the non-perturbative treatments for bound-state systems, as will be discussed in Secs.~\ref{subsec:atomic-system} and \ref{subsec:molecular-system}. Finally we note that if we consider a many-particle systems, i.e., when $\omega_d$ becomes large due to collective coupling and the matter system can span a large momentum scale, the above consideration shows that one cannot expect a reasonable approximation of the exact NRQED wave function by perturbation theory. Moreover, as can be seen from Eq.~\eqref{eq:renormalized-mass-derived}, also the number of particles within the effective mode volume affects the renormalized mass of the particles~\cite{rokaj2020}. Nevertheless, perturbative considerations with the appropriate cutoffs can be in qualitative agreement with non-perturbative results for specific observables such as ground-state energies.

Let us finally remark on the importance of the lower (infrared) cutoff in the numerically exact calculations and the consistency between light and matter. As is clear from the gauge coupling prescription, the fields are directly related to the matter wave functions~\cite{greiner1996,ruggenthaler2023}. So the matter grid determines which modes are possible. While we here have considered free particles, and the sizes of $L$ and $A$ are somewhat arbitrary, we aim at considering bound-state systems, for which the long-wavelength approximation is designed for. So the size of the simulation box is chosen such that all the relevant observables for the bound state are well converged. The free-space case is numerically very instructive to understand that a mismatch between the light and matter basis sets leads to non-physical results and that a non-perturbative study needs to be performed to have a consistent description of the coupled system at different light-matter coupling strengths. For instance, allowing for modes that are much smaller in energy than the minimal momentum eigenstate of matter, results in a wrong dispersion relation that becomes flat (see App.~\ref{sup:HEGlengthgauge} for an example). This again shows the importance of length scales in QED {in numerical calculations}~\cite{svendsen2023theory}, even if in the mathematically exact theory no divergence is found for soft photons~\cite{spohn2004}. Neglecting this important numerical detail would necessarily lead to erroneous results for the much more complex bound-state case, and hence checking the free-space situation first proves also paramount to generate reliable data.

\section{ The bound matter system coupled to the electromagnetic continuum}\label{sec:bound-matter}

The previous section considered the case of a single free charged particle interacting with the quantum fluctuations of a discretized continuum of isotropic photonic modes. We will now consider the case of a bound matter system interacting with the same discretized continuum of modes and investigate some effects the photon modes have on several physical properties of the coupled system. For our investigation of the bound system, we chose to work with the length form of the Pauli-Fierz Hamiltonian given by Eq.~(\ref{eq:length-gauge-hamiltonian}). A practical advantage of the length gauge Hamiltonian is that for a real-space description of the bound matter system, the spectrum converges faster for a basis of simple tensor products of photon displacement and matter states as opposed to the velocity gauge~\cite{han2010,bandrauk2013}. It is crucial to mention that the relation between the bare mass $m$ and the observable mass $m_e$ is non-perturbatively the same in both gauges. The energy dispersion of the free electron in the length-gauge has exactly the same form as the one obtained in the velocity gauge. We demonstrate this fact for the single-particle case in App.~\ref{sup:HEGlengthgauge}. At this point we would like to emphasize that to obtain the free-particle dispersion in the length gauge, the dipole self-energy is absolutely crucial. Without the dipole self-energy there is no translationally invariant direction in the electron-photon configuration space, i.e., translational invariance is broken~\cite{rokaj2017}. As a consequence, the free particle energy dispersion cannot be obtained non-perturbatively without the dipole self-energy. This makes evident the importance of the dipole self-energy for the mass-renormalization procedure.

To obtain physical observables of the coupled light-matter system, we solve the stationary eigenvalue problem $\hat{H}|\Psi_{n}\rangle = E_{n}|\Psi_{n}\rangle$ of the Pauli-Fierz Hamiltonian of Eq.~\eqref{eq:length-gauge-hamiltonian} and the matter-only QM setting of Eq.~\eqref{eq:matter-hamiltonian} numerically exactly and compare the results. In the {renormalized} setting for the bound systems, the contributions due to the interaction with a discretized continuum is accounted for by using the observable/renormalized mass obtained in Tab.~\eqref{tab:physical-masses} for the different couplings. We will consider two examples of one-dimensional model systems interacting with the electromagnetic continuum: the first being an atomic system consisting of a single bound electron, and the second a molecular hydrogen model of two interacting electrons and nuclei with soft-Coulomb potentials.

\subsection{The atomic light-matter system}
\label{subsec:atomic-system}

\begin{figure}
\includegraphics[width=1.0\columnwidth]{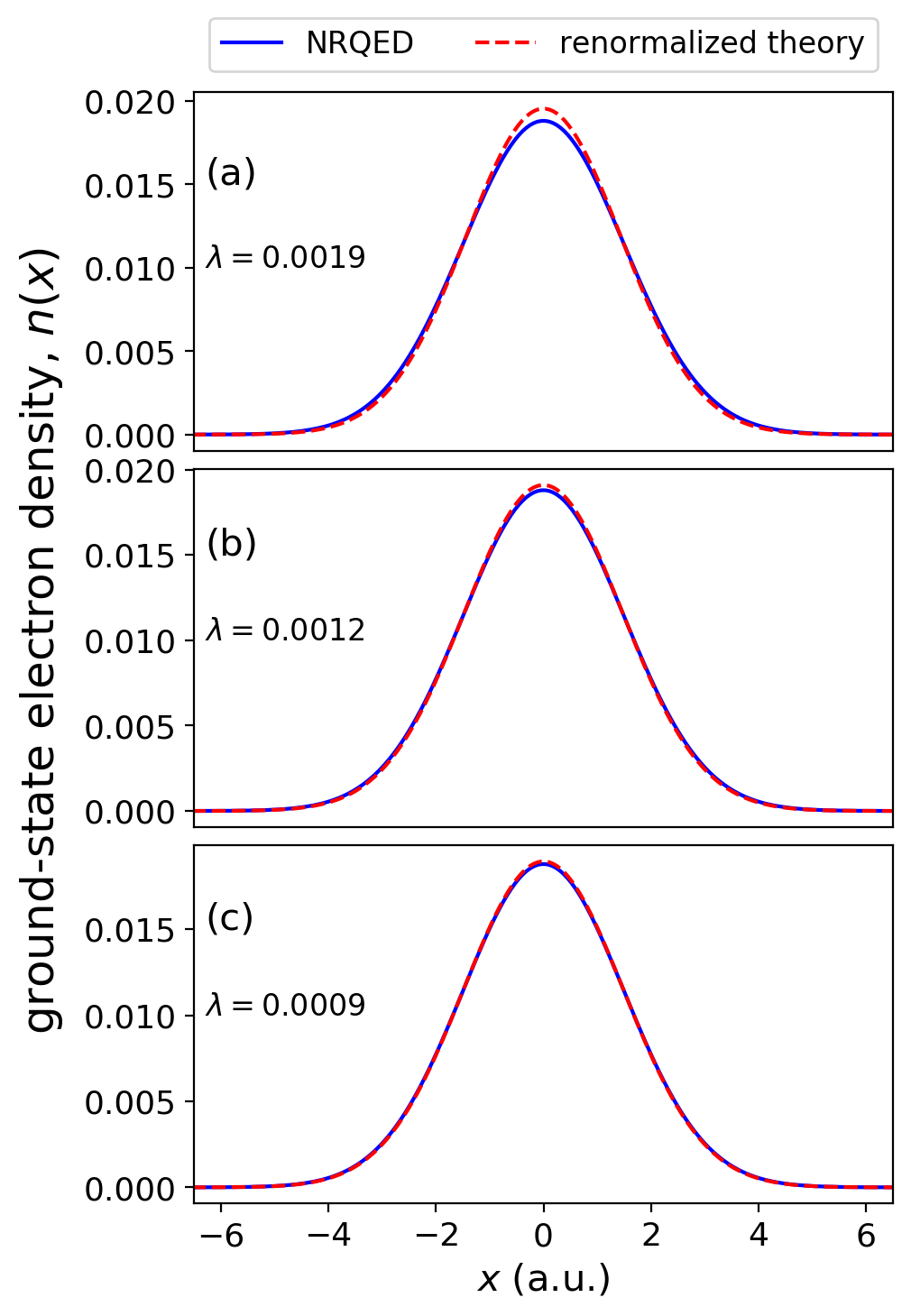}
\caption{A comparison of the electronic ground-state density between the NRQED and {the free-space renormalized} settings for a bound electron coupled to 200 photon modes. Panels (a-c) show the comparison for different coupling strengths $\lambda$ where the {renormalized} results deviate from NRQED due to how the bound system interacts with the discretized photonic continuum.}
\label{fig:ground-state-densities}
\end{figure}

In what follows we consider an atomic model of a single electron bound in the P\"{o}schl-Teller potential~\cite{Poschl1933}
\begin{eqnarray}
     v(x)=-\frac{e^{2}}{4\pi\epsilon_{0}}\frac{v_{0}}{\cosh^{2}(k_{0}x)}
\end{eqnarray}
where $v_{0}$ and $k_{0}$ are parameters that control the depth of the P\"{o}schl-Teller potential~\cite{Poschl1933}. {We note that the P\"oschl-Teller potential arises due to longitudinal interaction between a charged nucleus and a single charged electron and hence takes into account non-perturbatively the longitudinal photon interaction.} An important quantity of a matter system is the ground-state density, as it describes the localization properties of matter. {By localization of a bound system, we are referring to the exponential localization of the the probability distribution of finding a particle in a particular region trapped by an attractive longitudinal potential.} We investigate this property for the case of the one-dimensional atomic model (see App.~\ref{sup:1D-atomic-hydrogen} for the details of the model) coupled to the electromagnetic continuum. The atomic system interacts with the discretized continuum of photon modes discussed in Sec.~\ref{sec:free-electron}. For this setting of the coupled light-matter system, we compute the ground-state electron density, which gives the probability of finding an electron at position $x$, and make a comparison between the NRQED and {the free-space renormalized} settings. In contrast to the free-space dispersion, for which the mass-renormalization procedure was designed, we find that the observable-mass approximation does not lead to a quantitative agreement as illustrated in Fig.~(\ref{fig:ground-state-densities}) for the different light-matter coupling strengths. Instead, we find that the results from {the free-space renormalized theory} deviates from NRQED as the system in its ground-state becomes more bound as indicated by the increased amplitude and shrinking of the width of the density profile. In this setting of a bound system interacting with a continuum, it is interesting to find that the {usual free-space} mass-renormalization procedure does not agree with the NRQED results. 

\begin{figure}[bth]
\includegraphics[width=1.0\columnwidth]{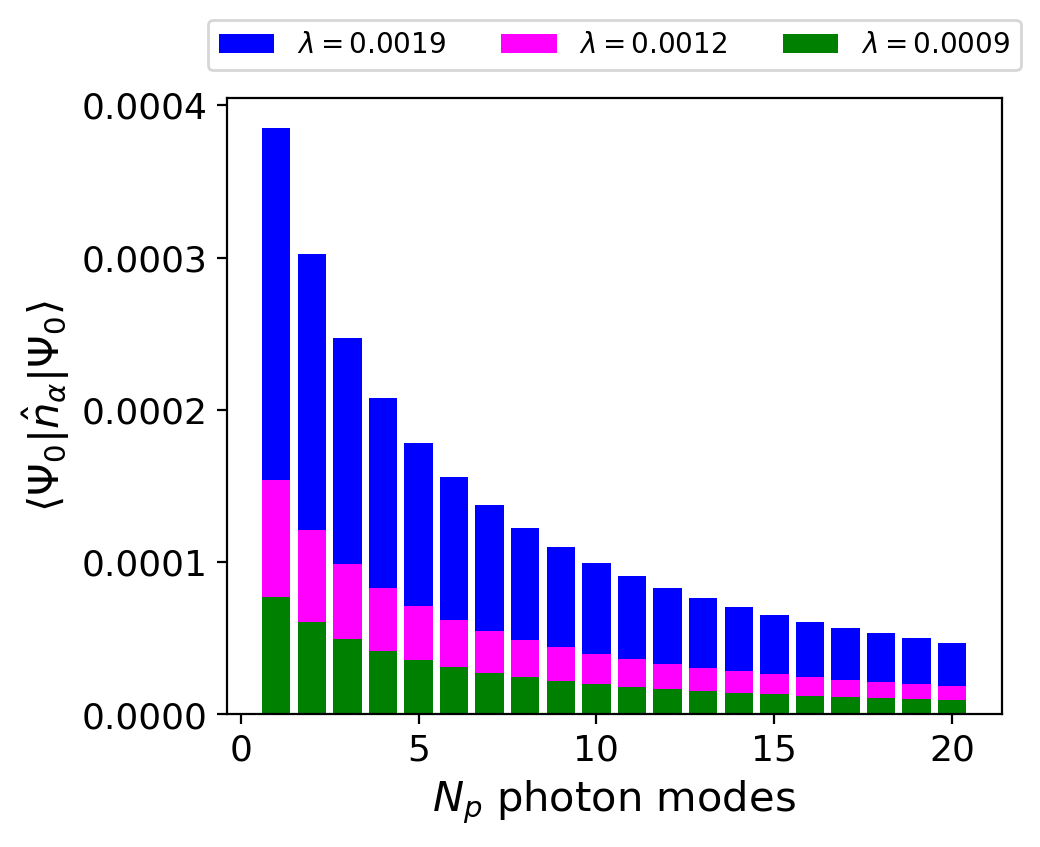}
\caption{The photon occupation of the electron-photon ground-state for the  atomic bound system coupled to a discretized continuum sampled with $N_{p}=200$ photon modes. Only the occupations of the first 20 low lying photon modes are shown where the lowest modes couple strongly and the occupation increases with the coupling strength. Each bar represents the contribution of each photon mode to the photon occupation.}
\label{fig:pt-occup-per-mode}
\end{figure}

The reason for this deviation is that by introducing a binding potential the localized electronic states do not couple equally to all the modes of the discretized electromagnetic continuum as opposed to the free-space case studied above. Since the energy dispersion of Fig.~\ref{fig:free-electron-M-modes} is actually an excited state property, it probes a larger part of the photonic continuum of modes. Hence, the resulting observable mass obtained from Eq.~\eqref{eq:renormalized-mass} (see values in Tab.~(\ref{tab:physical-masses})) includes a large contribution from the high-lying modes of the electromagnetic continuum. Therefore, using this renormalized mass in Eq.~\eqref{eq:matter-hamiltonian} leads to the deviations seen in Fig.~\ref{fig:ground-state-densities}. This is elucidated clearly in the second point where we illustrate how the different photon modes interact with the atomic system by computing the mean photon occupation per photon mode defined in the velocity gauge as $n_{\alpha} = \langle \Psi_{0}|\hat{n}_{\alpha}|\Psi_{0}\rangle$ where $\hat{n}_{\alpha}=\hat{a}_{\alpha}^{\dagger}\hat{a}_{\alpha}$ and $|\Psi_{0}\rangle$ is the correlated electron-photon ground-state. For different light-matter couplings $\lambda$, we show in Fig.~\ref{fig:pt-occup-per-mode} the mean photon occupation for the lowest lying 20 of the 200 photon modes. Clearly, the lower lying photon modes have more photon occupation as they interact more with the atomic system when compared to the high-lying modes. Also, we find that the stronger the coupling $\lambda$ the higher the photon occupation and the decreasing trend of photon occupation for higher lying photon frequencies applies for the different $\lambda$'s. From these results we can deduce that ground-state properties will saturate with increasing photon modes with higher frequencies (i.e., increasing photonic cutoff).   

\begin{figure}[bth]
\includegraphics[width=1.0\columnwidth]{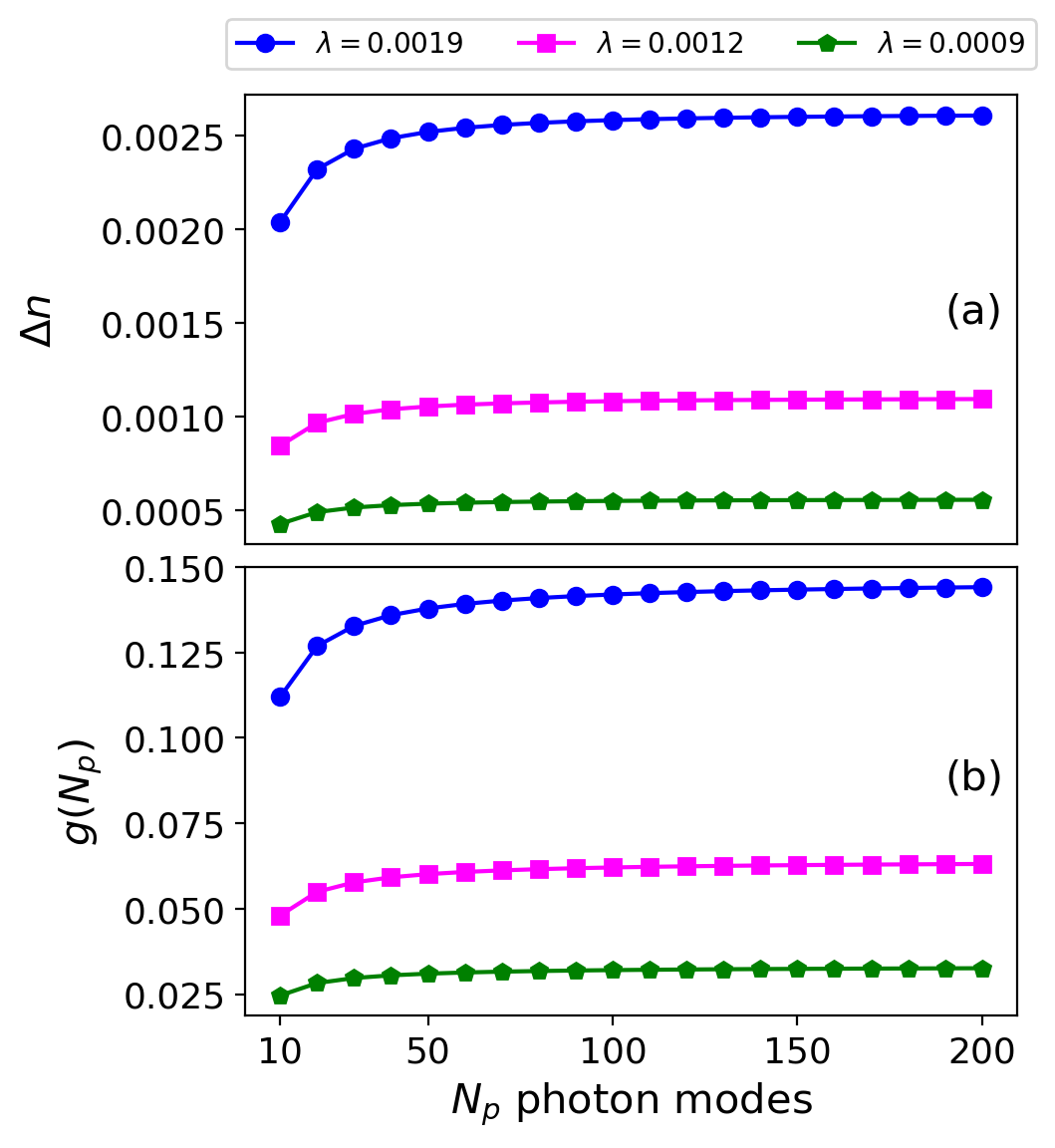}
\caption{(a) The integrated ground-state density difference $\Delta n$ between the NRQED and {the free-space renormalized} settings when coupled to different photon modes in the range 10 to 200 and increasing coupling $\lambda$. In contrast to the free-space case we see a convergence of the results when increasing the number of modes. (b) The dependence of the multi-mode coupling on the number of photon modes for different light-matter coupling.}
\label{fig:pt-density-per-mode}
\end{figure}

To demonstrate this, we compute the integrated ground-state electron density difference defined as $\Delta n = \int dx |n(x) - n'(x)|$ where $n(x)$ and $n'(x)$ are the densities of two different settings. Since NRQED is the reference result, we compute this quantity within this theory and compare then to {the renormalized theory} for increasing number of modes (increasing cutoff) from 10 to 200 modes in steps of 10 modes and increasing coupling $\lambda$. This comparison is shown in Fig.~\ref{fig:pt-density-per-mode}a where we find that ``NRQED$-${renormalized}" (density of NRQED {subtracted from the renormalized theory}) saturate for increasing photon modes for the different couplings. From the results of Fig.~ \ref{fig:pt-density-per-mode}a, we also infer that the atomic system becomes more bound (increased electronic localization) for increasing photon modes (photonic cutoff energy). A conclusion that can be drawn from the results of Figs.~\ref{fig:pt-occup-per-mode} and \ref{fig:pt-density-per-mode}a is that for the bound system not all photon modes are equally important since the effect of coupling to the ground-state becomes smaller for higher photon frequencies. This implies that the bound system saturates faster than the free particle as a function of the number of modes $N_{p}$. {It is interesting to highlight that the multi-mode coupling $g(N_{p})$ has a similar dependence on the number of photon modes as the integrated ground-state density up to a multiplicative prefactor as shown in Fig.~\ref{fig:pt-density-per-mode}b.} We would like to mention that the enhanced localization as a result of mass renormalization, has been reported even with a single cavity mode for a many-particle system in a harmonic potential~\cite{Rokaj2023collective}. In this case the localization phenomenon was significantly enhanced due to the collective coupling of the system and cavity-mediated interactions. A further important point to make here is that, as opposed to the free-space case, where Eq.~\eqref{eq:renormalized-mass-derived} shows that a finite cutoff/regularization needs to be kept, for ground-states, even in the long-wavelength approximation NRQED might become largely cutoff-independent for a fixed bare-mass value.  Finally we note that an exponentially suppressed mode occupation for higher frequencies reflects the origin of quantum physics, where quantized photon modes were introduced to overcome the Rayleigh–Jeans ultraviolet catastrophe.\\

\begin{figure}[bth]
\includegraphics[width=1.0\columnwidth]{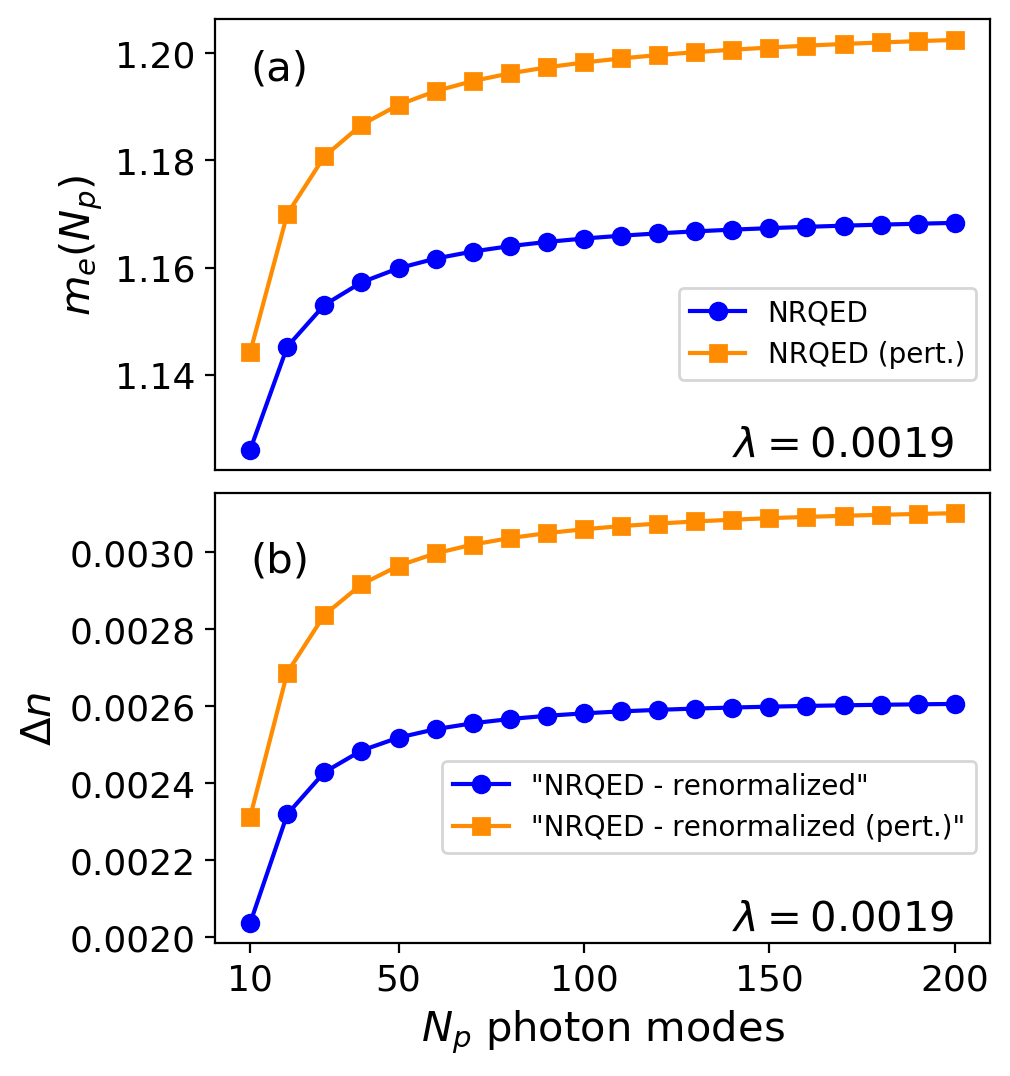}
\caption{(a) Comparison between the renormalized mass of the exact NRQED and that using perturbation theory. (b) The integrated ground-state density difference $\Delta n$ between NRQED and {the renormalized theory}, and NRQED and {the perturbatively-renormalized theory}.}
\label{fig:pt-density-per-mode-comp}
\end{figure}

Before moving on, we provide results for the ground-state properties of the atomic light-matter system using perturbation theory. In this approach, using Eq.~\eqref{eq:free-electron-photon-spectrum-pert} we determine the renormalized electron mass as noted from Eq.~\eqref{eq:physical-mass} and given in Tab.~(\ref{tab:physical-masses}), which is then used in the {renormalized theory}. We show in Fig.~\ref{fig:pt-density-per-mode-comp}a how the renormalized mass obtained from perturbation theory deviates from the exact NRQED case. Perturbation theory will have an increasing deviation for higher-lying photon modes. Furthermore, to demonstrate how the perturbative treatment differs from the non-perturbative results, we compute the integrated ground-state electron density difference between NRQED and the {perturbatively-renormalized} results and make a comparison with the non-perturbative results. This result is shown in Fig.~\ref{fig:pt-density-per-mode-comp}b where we find a similar behavior between exact NRQED and perturbation theory stemming from the respective renormalized masses employed. Based on these findings, we obtain the important result that perturbation theory deviates more from the exact results as the number of photon modes increases.

\subsubsection{Impact of Mass Renormalization on excited-state properties}
\label{subsubsec:atomic-excited-state}

\begin{figure}[bth]
\includegraphics[width=1.0\columnwidth]{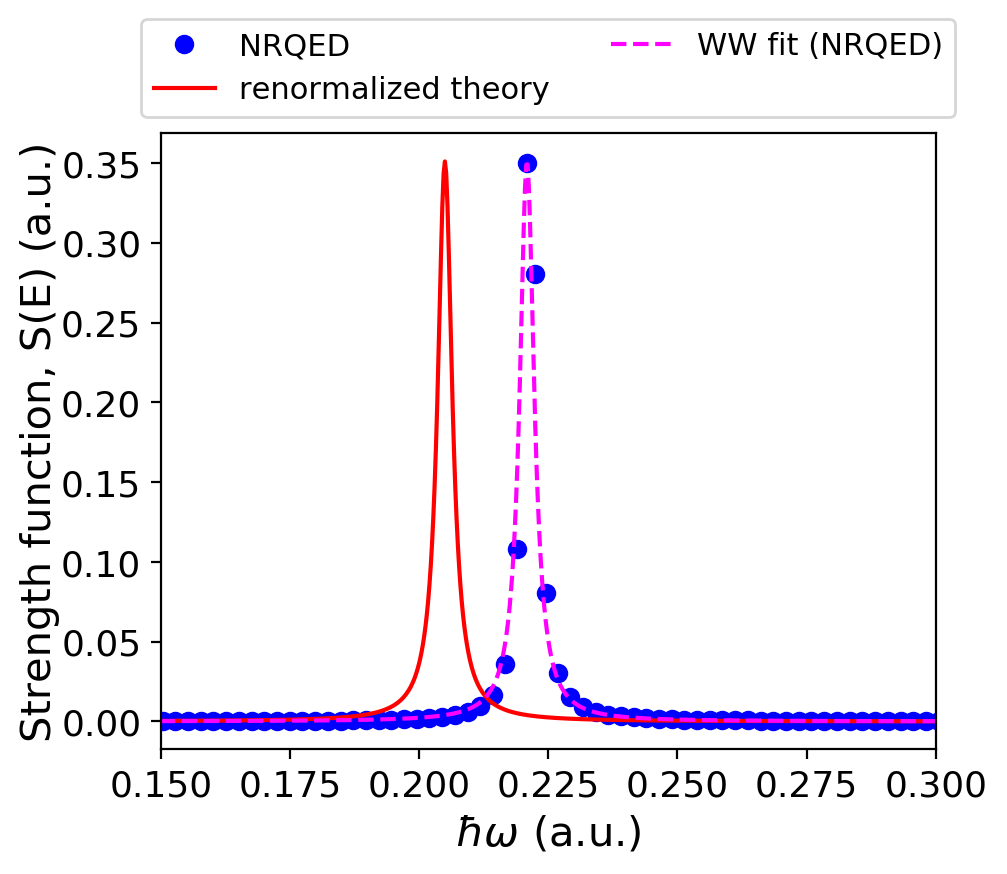}
\caption{The absorption spectrum of the NRQED setting without applying an artificial broadening and the corresponding fitted spectrum with Wigner-Weisskopf (WW) theory (magenta) with a Lorentzian broadening of $\hbar\gamma=0.0034$~a.u. The same broadening and WW theory is employed to obtain the absorption spectrum of the {free-space renormalized} setting (red).}
\label{fig:abs-cross-section}
\end{figure}

We have so far considered only ground-state properties for the atomic system interacting with the discretized electromagnetic continuum. Now, we focus on excited-state properties of this coupled system. One common quantity which is normally computed is the absorption spectrum of the system which we determined here by computing the dipole strength function $S(\omega)=\sum_{n}2\omega_{n}|\langle\Psi_{0}|\hat{x}|\Psi_{n}\rangle|^{2}\delta(\omega-\omega_{n})$ where $\hat{x}$ is the electronic dipole operator of the one-dimensional atomic system. For this quantity, we make a comparison for the different settings as shown in Fig.~\ref{fig:abs-cross-section} where we employ Wigner-Weisskopf theory to fit the NRQED results and subsequently obtain the spectrum of the {renormalized} setting. Similar to the ground-state results, we find that {free-space renormalized theory} deviates from NRQED in peak position. The reason for this can be partly attributed to how the discretized continuum of modes interact with the atomic system and affect ground-state properties such that a transition from the ground-state to the first excited state leads to this deviation. We expect that for higher-lying excitations in the absorption spectrum, the NRQED and {renormalized} settings should agree since the excited states become more delocalized and should probe a large part of the continuum as in the free-electron case discussed above. A noticeable difference is that the absorption peak of {the free-space renormalized theory} is red-shifted in the spectrum relative to NRQED. From the analytic expression of the energies of the atomic system given in Eq.~(\ref{eq:analytic-spectrum}), we deduce that for a larger (i.e. the observable) mass, the energies become more negative (strongly bound) which causes the red shift relative to the NRQED peak. In passing, we note that for a more dense sampling of the discretized continuum of modes as done in Refs.~\cite{flick2019,welakuh2022,welakuh2022a}, we will obtain a smooth Lorentzian profile for the NRQED case that naturally occurs due to the continuum of modes.

\subsection{The molecular light-matter system}
\label{subsec:molecular-system}

\begin{figure}[bth]
\includegraphics[width=1.0\columnwidth]{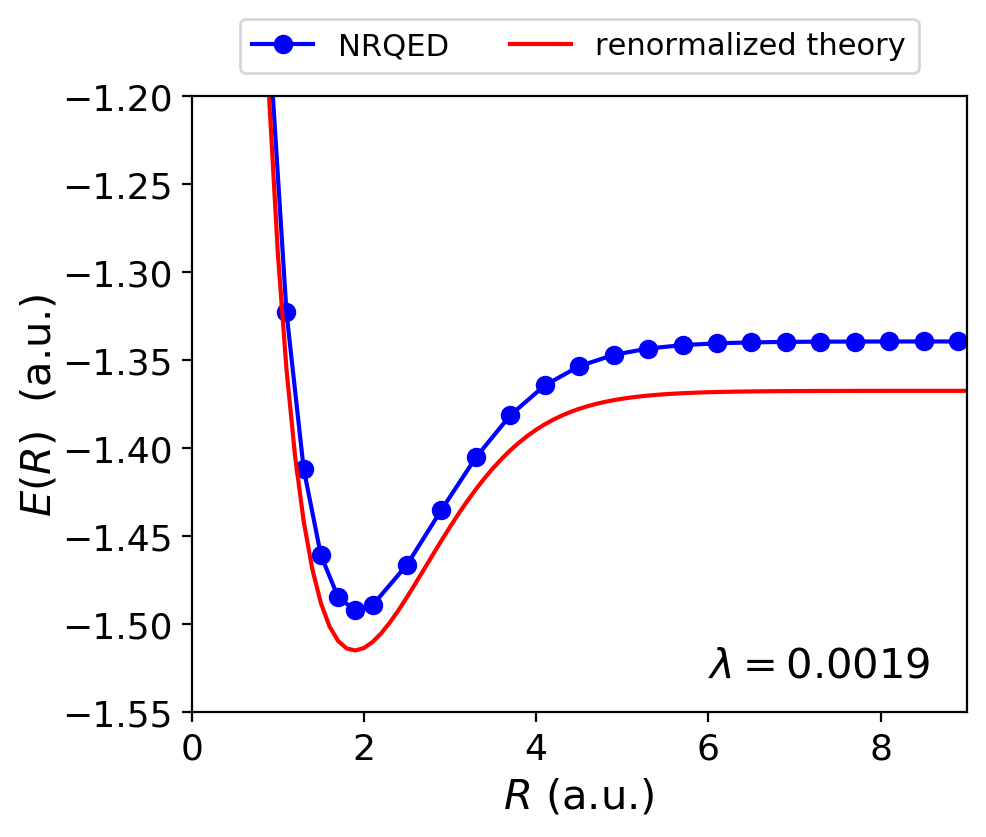}
\caption{A comparison of the ground-state PoPESs showing a deviation between the {free-space renormalized} and NRQED results. The PoPES for NRQED is the case when coupled to 15 photon modes and {renormalized theory} uses the observable mass due to the same photonic density of states. }
\label{fig:BOPESs-QED-vs-renorm}
\end{figure}

We now investigate how molecular properties are affected when a molecular system interacts with the discretized continuum of photon modes. Our example considers the model for the  H$_{2}$ molecule where the motion of all particles is restricted to one spatial dimension and the center-of-mass motion of the molecule can be separated off {similar to the atomic case discussed above}~\cite{lively2021,albareda2021,kreibich2001}. The relevant coordinates of this model are the internuclear separation, $R$, and the two electronic coordinates, $x_{1}$ and $x_{2}$. The Hamiltonian of the molecular model is 
\begin{align}
\hat{H}_{\text{mol}} &= - \frac{1}{2\mu_{n}}\frac{\partial^{2} }{\partial R^{2}} + \frac{1}{R} + \frac{1}{\sqrt{(x_{1}-x_{2})^{2} + a_{\text{ee}}}} \nonumber \\
& \quad + \sum_{i=1}^{2} \left(- \frac{1}{2\mu_{\text{e}}}\frac{\partial^{2} }{\partial x_{i}^{2}} -\frac{1}{\sqrt{(x_{i}-R/2)^{2} + a_{\text{en}}}} \right .\\
&\qquad\qquad\quad \left .  - \frac{1}{\sqrt{(x_{i}+R/2)^{2} + a_{\text{en}}}}\right) , \nonumber
\label{eq:1D-H2-hamiltonian}
\end{align}
where $\mu_{\text{e}}=2M_{n}/(2M_{n}+1)$ and $\mu_{n}=M_{n}/2$ are the reduced electronic and nuclei masses, respectively. Further details on how we treat the one-dimensional model of a hydrogen molecule is provided in App.~\ref{sup:1D-molecular-hydrogen}.

For the calculations of the molecular light-matter system, we couple the molecule to the lowest 15 photon modes since they are the most important for bound systems as discussed above. A common and widely studied property of a molecule is its potential energy surface (PES) which describes the relationship between the molecular geometry, for example, the relative positions of the participating atoms, and the molecular energy. For the case of coupled light-matter systems we have similar objects. Since we have three natural subsystems, i.e., nuclei, electrons and photons, we can perform the Born-Huang expansion that underlies the PES concept in different ways~\cite{flick2017b, schaefer2018, ruggenthaler2023}. In our case, where the frequency range is chosen to affect the electronic degrees of freedom (as we show in the App.~\ref{sup:1D-molecular-hydrogen}, the basic frequency of the nuclear degrees of freedom is $\omega_{e} \approx 0.02 $ a.u. which is within the lower frequencies of the sampled continuum), we can use a grouping of the photonic degrees of freedom with the electronic ones. This leads to polaritonic PES (PoPES)~\cite{feist2015}, where the nuclei (in our case indicated by the internuclear separation $R$) `feel' the photonic continuum of modes via the changes in the PoPES.

We now show in Fig.~\ref{fig:BOPESs-QED-vs-renorm} the ground-state PoPES for the different settings. This result is similar to the atomic light-matter system discussed above, where the lower-lying modes of the continuum couple strongly compared to the higher-lying modes which causes the deviation when the calculated renormalized mass (in Tab.~(\ref{tab:physical-masses})) is used in the {renormalized theory}. To support this, we show in Fig.~\ref{fig:gs-density-QM-vs-bQM} the ground-state density of NRQED and {of the free-space renormalized theory}, where we find that the {renormalized} result shows that the molecular system at the equilibrium position becomes more bound when compared to NRQED, similar to the atomic light-matter results in Fig.~\ref{fig:ground-state-densities}a. We note that we have removed the vacuum contribution of the zero-point energy due to the 15 photon modes from the PoPES of NRQED. That means, we have normal-ordered and discarded an overall constant energy contribution.

\begin{figure}[bth]
\includegraphics[width=1.0\columnwidth]{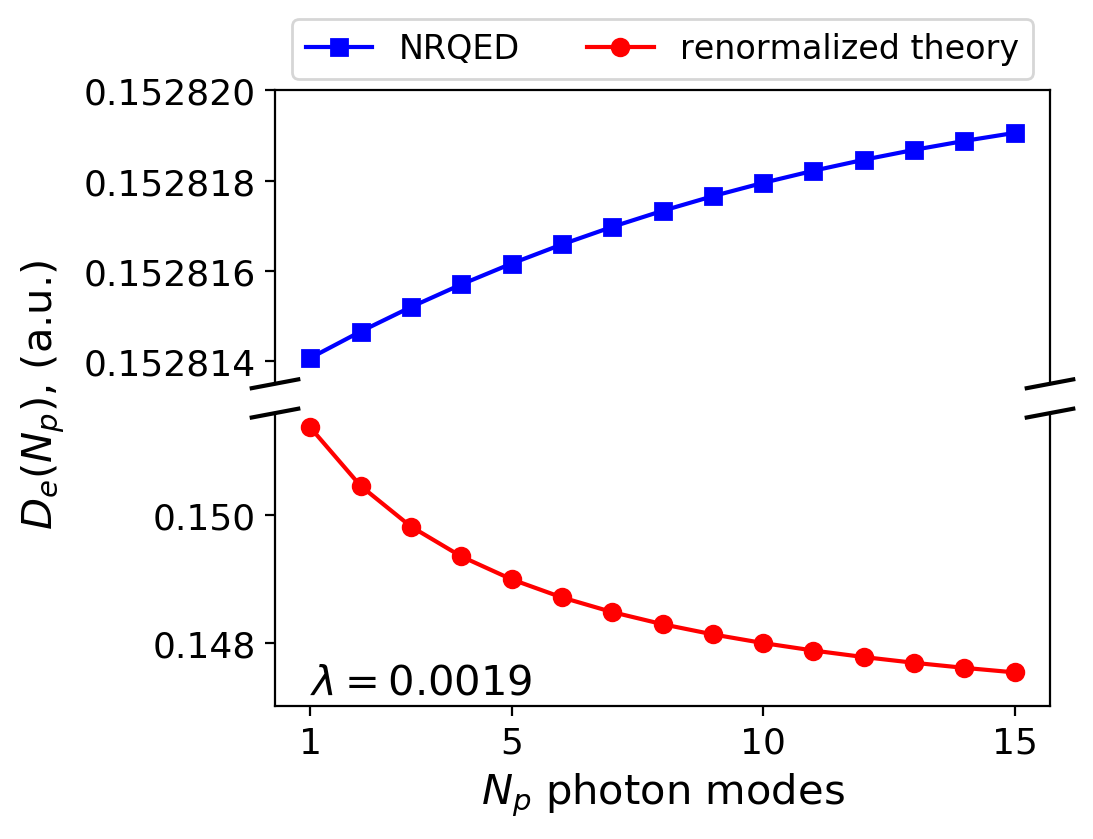}
\caption{The photon-mode-dependent dissociation energy plotted as function of photon modes for the NRQED and {the free-space renormalized theory}. The NRQED setting indicates that it is increasingly difficult to break a bond for increasing $N_{p}$ while {the renornmalized theory} shows the opposite behavior.}
\label{fig:BOPESs-QED-vs-renorm-diss-ener}
\end{figure}

Another important information that can be obtained from Fig.~\ref{fig:BOPESs-QED-vs-renorm} is the bond dissociation energy $D_{e}$ of the molecule. The results on how the dissociation energy of the H$_{2}$ molecule changes with increasing photonic energy cutoff (number of photon modes) is shown in Fig.~\ref{fig:BOPESs-QED-vs-renorm-diss-ener}. For up to 15 photon modes, the photon-mode-dependent dissociation energy for NRQED and {the free-space renormalized theory} has an opposite behavior. For NRQED the dissociation energy increases with increasing energy cutoff which implies that it is more difficult to break a chemical bond when the molecule is made to interact with the electromagnetic continuum. This is in line with the enhanced binding of the minimal-coupling Pauli-Fierz Hamiltonian due to trapping of virtual photons~\cite{hiroshima2012enhanced, spohn2004}. The trapping of virtual photons is nicely demonstrated in Fig.~\ref{fig:pt-occup-per-mode}. That the renormalized theory shows the opposite trend is due to the fact that it spuriously assumes that all modes couple equally strong irrespective of the localization of the electronic system, similar to the atomic case. In other words, if we would like to recover the NRQED results we would need to have also an $R$ dependence in the renormalized mass, where only for large $R$ we would approach the free-space value.

\begin{figure}[bth]
\includegraphics[width=1.0\columnwidth]{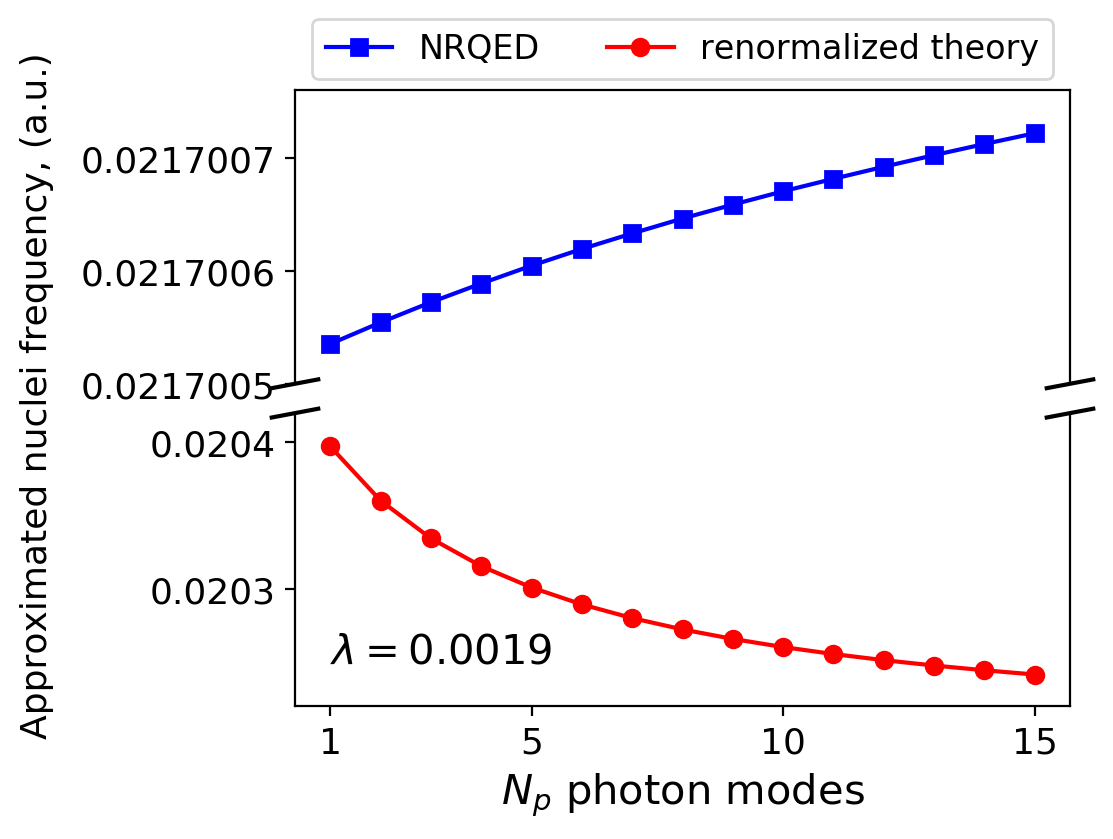}
\caption{A comparison of the harmonic-approximated frequency of the ground-state PoPES between NRQED and {the free-space renormalized theory} as a function of the number of photon modes $N_{p}$. The approximated harmonic frequency for $N_{p}=0$ is $\omega_{e}=0.020455$~a.u.}
\label{fig:approximated-harmonic-frequency}
\end{figure}

As we have seen, the PoPES changes due to the interaction with the photon modes. To quantify the effect of the modes on the nuclear degrees further we next consider the change in vibrational frequencies in the $H_2$ molecule. Since we have chosen a Born-Huang grouping of the electrons with the photons, the effect of the many modes is mediated via the changes in the PoPES. We note that for free interacting protons coupled to the electromagnetic continuum, we obtain an analogous dispersion energy as in Eq.~\eqref{eq:free-electron-photon-spectrum} with a diamagnetic frequency that is dependent on the nuclear charge. From the energy dispersion the renormalized proton mass can be obtained and with this we can investigate how the nuclear degrees are influenced due to coupling to the electromagnetic continuum in a {renormalized} setting. In Fig.~\ref{fig:approximated-harmonic-frequency} we show the results of the harmonically approximated vibrational frequency dependence on the sampled photon continuum (see App.~\ref{sup:morse-fit-PES} for details).  We find for NRQED that the approximate harmonic vibrational frequency increases with the number of photon modes indicating that the nuclear degrees of the ground-state PoPES becomes more bound while the {free-space renormalized theory} shows the opposite behavior. The behavior of the approximate harmonic frequency is reminiscent of the dissociation energy in Fig.~\ref{fig:BOPESs-QED-vs-renorm-diss-ener} since it is proportional to the square-root of $D_{e}$. We can thus conclude that the nuclear degrees of freedom are influenced in a similar way to the electronic degrees where only the lower-lying photon modes play a significant role.

In addition, the fact that the PES of the $H_2$ molecule is modified indicates that the photon field modifies the long-range interactions between the atomic dimer. The intuition that the mediated forces are of long-range nature is due to the fact that if we fit the PESs, with and without coupling to the photon field, we find that the defining parameters ($D_e, a, \textrm{and}\;\omega_e$) of the Morse model are modified when the molecule is coupled to light. The Morse model effectively describes the long-range interactions between the pair of atoms, which are responsible for the formation of the molecule. Thus, it becomes evident that the photon field has an impact on these long-range interactions, which is in line with recent experimental results~\cite{ebbesen2023direct}. 

\subsection{Cavity-modifications of the ground-state}
\label{subsec:cavity-modifications}

In this section we focus on how enhancing the coupling between the bound matter system and the electromagnetic continuum can lead to the modification of ground-state properties. There are several methods by which the coupling to the electromagnetic continuum can be enhanced. For instance, to enhance the coupling of a single atom or molecule usually micro- and nanocavities are employed~\cite{skolnick1998strong,raimond2001manipulating,vahala2003optical,hugall2018plasmonic}, while to enhance the collective coupling often Fabry-P\'erot cavities are used~\cite{garcia-vidal2021}. Here, we distinguish two settings for the coupling of matter with vacuum fluctuations. The first setting is the reference ``free-space'' case where the discretized continuum is weakly coupled to matter and we chose $\lambda=0.0009$ to designate the free-space coupling. In the second setting, the discretized continuum is made to strongly interact with matter by decreasing the mode volume for the relevant frequency ranges to enhance the coupling. In this cavity setup, we enhance the coupling to the discretized continuum by increasing the coupling parameter $\lambda=0.0012, 0.0019$. As we learned from Sec.~\ref{sec:free-electron}, we need to have the matter and the photonic degrees of freedom to be consistent. We therefore only consider the photonic density of states in the relevant frequency range, where there are matter states that can be affected by the photons. We do not consider how the photonic modes are changed outside of this frequency range, from where the extra density of states is taken from. The specifics of the photonic environment are not further discussed here, but engineering the photonic modes can be done in an ab initio setting via, e.g., macroscopic QED~\cite{svendsen2024ab}. We will investigate the properties of the coupled system in the following only in the NRQED setting.

\begin{figure}[bth]
\includegraphics[width=1.0\columnwidth]{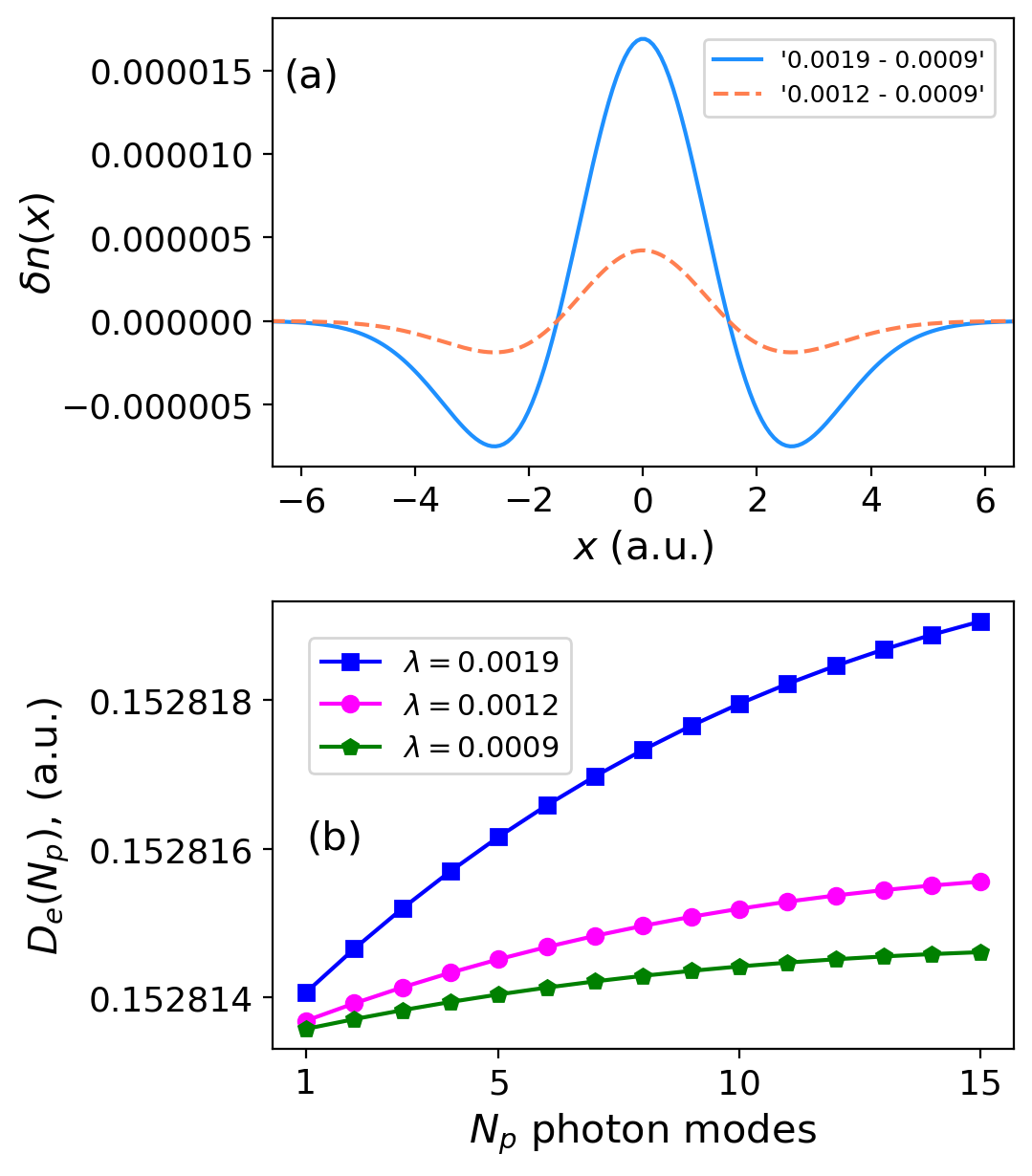}
\caption{(a) The ground-state density difference $\delta n(x)$ between the cavity and free space case for two cases (blue: ``$0.0019 - 0.0009$") and (orange: ``$0.0012 - 0.0009$"). (b) A comparison of the photon-mode-dependnet dissociation energy for different light-matter coupling. }
\label{fig:gs-density-BOPES-NRQED-QM}
\end{figure}

In Fig.~\ref{fig:gs-density-BOPES-NRQED-QM}, we show the results of the atomic and molecular light-matter systems for the free-space and the cavity settings. For the atomic system in Fig.~\ref{fig:gs-density-BOPES-NRQED-QM}a, we compute the difference between the ground-state density $\delta n(x)$ of the free space case ($\lambda=0.0009$) and when we enhance the coupling to the discretized continuum with cavity with couplings $\lambda=0.0012, 0.0019$. Although relatively small, we find that there are cavity-induced modifications of the ground-state density (i.e. atomic system is more bound) when we change the photonic continuum using a cavity. For the molecular light-matter in Fig.~\ref{fig:gs-density-BOPES-NRQED-QM}b, we show the photon-mode-dependent dissociation energy of the ground-state PoPES where we find that it becomes more difficult to break a chemical bond when the cavity mode enhances the coupling to discretized continuum. On the other hand, if we would change the photonic environment to have less photonic density of modes in the relevant frequency range, we would actually observe the opposite effect. These results highlight that ground-state properties of bound systems can be modified when the coupling to the photonic continuum is changed, for example, using an optical cavity. The specific effect will depend on the properties of the cavity, i.e., how it re-arranges the mode strengths locally, and on the matter system, i.e., in which range of energies the matter wave function has occupations~\cite{svendsen2023theory}.

\section{Summary, Conclusion and Outlook}
\label{sec:conclusion-outlook}

We have investigated non-perturbatively how the coupling to a (discretized) continuum of modes leads to mass-renormalization effects in ab initio QED. Starting with free charged particles interacting with a photonic continuum of modes, we demonstrated the free-space mass-renormalization procedure and highlighted how it connects two levels of theory (NRQED {and the free-space renormalized theory}) by the energy dispersion. We showed the dependence of the renormalized mass on the amount of photon modes and the light-matter interaction strength. Our numerically exact, non-perturbative simulations highlighted the shortcomings of second-order perturbation theory. We found in accordance to intuition that the renormalized mass increases when the coupling is increased in the relevant energy range and decreases for deceasing coupling. In the non-perturbative mass renormalization case of bound systems coupled to light, we found that the NRQED and {renormalized} settings, which are connected via the free-space mass renormalization, do not agree for both its equilibrium and excited-state properties. This occurs because the bound system interacts differently with the electromagnetic continuum as opposed to the free-particle case. That is, out of the sampled discretized continuum only a few of the lowest-lying photon modes play a significant role when interacting with a bound system. 

{These results, which highlight that the interplay between the longitudinal and transverse light-matter interactions is non-trivial, also demonstrate that one should be cautious when using a multi-mode description of a photonic environment together with a free-space renormalized description of the charged particles. This is a very relevant and timely topic, since in ab-initio QED simulations routinely the free-space renormalized masses of the charged particles are used in conjunction with multiple modes of a cavity~\cite{ruggenthaler2023}. As has been highlighted in the literature~\cite{ruggenthaler2023, svendsen2023theory}, one at least needs to subtract the free-space continuum of modes from the cavity description. Yet, as we have demonstrated in this work, the interplay between the free-space renormalization and the longitudinal interaction between the charged particles can be highly non-trivial. Although these two effects are individually well-studied in their respective communities, their interplay clearly merits future research and is clearly important for polaritonic chemistry and cavity materials engineering.}

Naturally the question arises, whether {this non-trivial interplay} will also be found in real three-dimensional ab initio systems. The main difference when going to three dimensions will be the anisotropy of the renormalized masses. The cavity breaks the simple free-space symmetries and it will be interesting how these symmetry-breaking can influence real systems. It is further interesting to study how simple approximation schemes like lumping the effect of the continuum of modes into a few effective modes for ground-state properties behave~\cite{svendsen2023theory}. Such studies might allow to qualitatively study mass-renormalization effects also for realistic three-dimensional systems. The renormalization effects will also change if we go beyond the long-wavelength approximation. Although in free-space there is a fundamental difference between minimal-coupling and the long-wavelength approximation, i.e., dipole approximation is not fully non-perturbatively renormalizable yet for the full Pauli-Fierz Hamiltonian it might be possible similar to the Nelson model~\cite{spohn2004,hiroshima2005}, the saturation effect for bound states might point towards a very similar behavior of the long-wavelength and the minimal-coupling situation. Also, based on the success of the dipole approximation for bound systems, it seems reasonable to assume that in such cases the differences are usually small. Clearly, there are many cases where one expects stark differences, such as due to self-organization in a cavity or when large momenta are transferred between light and matter. Overall we believe, however, that the obtained model results are a very good indicator when similar effects will appear in realistic ab initio systems. Understanding these inherently non-perturbative effects could help us to get a further theoretical control-knob on the properties of matter in photonic environments. Besides this more practically relevant implications, the obtained theoretical insights could provide a different viewpoint, from a non-perturbative perspective, on the renormalization effects that show up in interacting quantum field theories. If QED phenomena can be described fully non-perturbatively, without resorting to infinite renormalization, it could support the idea by Van Hove~\cite{VANHOVE1952145} that divergences in quantum field theory are not a general property, but could be due to applying perturbation theory.



\section*{Acknowledgement}

We acknowledge enlightening discussions with Johannes Flick, Christian Eckhardt, Mark Kamper Svendsen and Heiko Appel. DW was supported by a grant from the Simons Foundation (Grant 839534, MET). We acknowledge support from the Max Planck-New York City Center for Non-Equilibrium Quantum Phenomena. The Flatiron Institute is a division of the Simons Foundation. MR and AR acknowledge support by the Cluster of Excellence “CUI: Advanced Imaging of Matter” of the Deutsche Forschungsgemeinschaft (DFG), EXC 2056, project ID 390715994 and the Grupos Consolidados (IT1453-22). VR acknowledges support from the NSF through a grant for ITAMP at Harvard University.


\appendix

\section{Numerical Details}
\label{sup:numerical-details}

We outline the numerical details to treat the coupled matter-photon system. First, for the matter Hamiltonian of the one-dimensional atomic system, we represent the single bound electron on a uniform real-space grid of $N_{x} = 3000$ grid points with grid spacing $\Delta x = 0.0707$ a.u. while applying an eighth-order finite-difference scheme for the momentum operator and Laplacian. Next, we perform an exact diagonalization of the Hamiltonian and obtain the spectrum of the system (converged eigen-energies $E_{i}$ and eigen-states $|\psi_{i}\rangle$). Now, using the completeness relation $\sum_{i=1}^{\infty}|\psi_{i}\rangle\langle\psi_{i}|=\hat{\mathbb{1}}$, the operators of the matter system can be expressed as~\cite{loudon2000}
\begin{align*}
\hat{H}_{\text{M}} &= \sum_{i=1}E_{i}|\psi_{i}\rangle\langle\psi_{i}|, \;\; \hat{\textbf{p}} = \sum_{i=1}\sum_{j=1} \langle\psi_{i}|\hat{\textbf{p}}|\psi_{j}\rangle |\psi_{i}\rangle\langle\psi_{j}|, \; \\ 
\hat{\textbf{R}} &= \sum_{i=1}\sum_{j=1} \langle\psi_{i}|\hat{\textbf{R}}|\psi_{j}\rangle |\psi_{i}\rangle\langle\psi_{j}|, \nonumber
\end{align*}
where the indices $i,j$ runs over the number of matter states considered. We consider $N_{s}=10$ lowest energy states for {the models of atomic and molecular hydrogen} to couple to the electromagnetic field. For the photonic subsystem, each photon mode is represented in a basis of Fock number states. {For the atomic light-matter system,} to be able to treat the discretized photonic continuum consisting of $N_{p} = 200$ photon modes numerically exact, we truncate the Fock space and consider only the vacuum state, the $N_{p}$ one-photon states, and the $(N_{p}^{2}
+ N_{p})/2$ two-photon states as in Ref.~\cite{flick2017}. This implies the dimension of the photonic continuum is $1+N_{p}+(N_{p}^{2}+ N_{p})/2=20301$. Coupling to $N_{s}=10$ lowest energy states of the atomic system give an atom-photon dimension of $10\times 20301=203010$. {We note that keeping up to the two-photon states for the different light-matter coupling strengths is sufficient to obtain numerical convergence. For example, the integrated ground-state density difference between the case including up to two-photon states and that where we keep only the vacuum state and the $N_{p}$ one-photon states is $\Delta n = 8.238 \times 10^{-10}$.}

For the model of the hydrogen molecule (H$_{2}$) in 1D, we used a grid $(0, 9]$ au for the internuclear separation with a uniform grid spacing $\Delta R = 0.1$ a.u. For the electron coordinates ($\hat{x}_{1}$ and $\hat{x}_{2}$),  we represent both electrons on a uniform real-space grid of $N_{x_{1}} = N_{x_{2}}= 200$ grid points  with grid spacing $\Delta x_{1} = \Delta x_{2} =  0.35$ a.u. {We perform exact numerical diagonalizations to obtain the spectrum and use only the $N_{s}=10$ lowest energy states for different $R$. We couple to the discrete photonic continuum as described above but for the molecular light-matter system, we include up to five-photon Fock states for each photon mode to obtain numerical convergence.}

\section{Perturbative and Exact Free Particle Dispersion and Continuum Behaviors}
\label{sup:nonperturbative-vs-perturbative}

\begin{figure}
\centering
\includegraphics[width=0.8\linewidth]{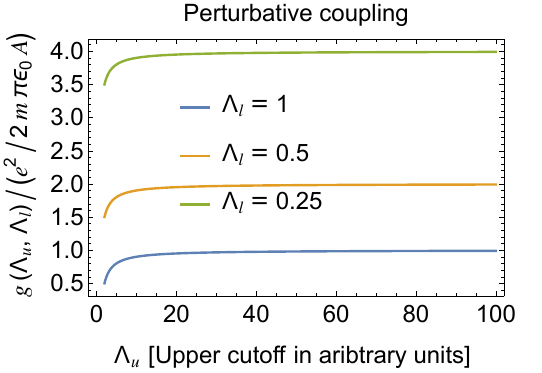}
\caption{Normalized perturbative light-matter coupling $g(\Lambda_u,\Lambda_l)$ for different values of the lower cutoff $\Lambda_l$. The coupling constant increases rapidly and asymptotically reaches a fixed value which is inversely proportional to the lower cutoff $\Lambda_l$. For $\Lambda_l \to 0$ the perturbative coupling diverges.}
\label{perturbative_coupling}
\end{figure}

In this section we compute the free particle dispersion in 1D perturbatively and we compare to the exact non-perturbative solution. Following Ref.~\cite{craig1998} the first non-trivial correction to the free particle dispersion in three dimensions is
\begin{eqnarray}
\Delta E=\frac{e^2}{m^2} \sum_{\mathbf{p}^{\prime}, \mathbf{k},\alpha}\frac{\hbar \mathbf{e}_{\alpha}(\mathbf{k})\cdot \mathbf{e}_{\alpha}(\mathbf{k})}{2\epsilon_0 c|\mathbf{k}| V} \frac{\langle \phi_{\mathbf{p}}|p_i|\phi_{\mathbf{p}^{\prime}}\rangle\langle \phi_{\mathbf{p}^{\prime}}|p_j|\phi_{\mathbf{p}}\rangle}{\mathbf{p}^2/2m-\mathbf{p^{\prime}}^2/2m -\hbar c|\mathbf{k}|}.\nonumber\\
\end{eqnarray}
In our effective one-dimensional model the polarization vectors are all parallel and the correction to the energy dispersion simplifies
\begin{eqnarray}
\Delta E=\frac{e^2}{m^2} \sum_{p^{\prime},k} \frac{\hbar}{2\epsilon_0 c |k| V} \frac{\langle \phi_{p}|p|\phi_{p^{\prime}}\rangle\langle \phi_{p^{\prime}}|p|\phi_{p}\rangle}{p^2/2m-{p^{\prime}}^2/2m -\hbar c|k|} .
\end{eqnarray}
Then, we use the property for the plane waves $\langle \phi_{p}|p|\phi_{p^{\prime}}\rangle=p\delta_{pp^{\prime}}$, we sum over $p^{\prime}$ and we find
\begin{equation}
\Delta E=-p^2\frac{e^2}{m^2c^2} \frac{1}{2\epsilon_0 V}\sum_{k}\frac{1}{k^2} \, . 
\end{equation}
In contrast to the main part, we here perform the summation over all photonic momenta $k$ by promoting the sum into an integral. For this purpose we write the mode volume as $V=AL$ and we have
\begin{equation}
\Delta E=-p^2\frac{e^2}{m^2c^2} \frac{1}{4\pi\epsilon_0 A} \int^{\Lambda_{u}}_{\Lambda_{l}} \frac{dk}{k^2} \, ,
\end{equation}
where $\Lambda_{u}$ and $\Lambda_{l}$ are the limits of integration. After the integration we find 
\begin{eqnarray}
\Delta E=-\frac{p^2}{2m}\frac{e^2}{2m c^2 \pi \epsilon_0A}\left(\frac{1}{\Lambda_{l}}-\frac{1}{\Lambda_{u}}\right)=-\frac{p^2}{2m}g(\Lambda_{u},\Lambda_{l}).\nonumber\\
\end{eqnarray}
From the expression of the multi-mode coupling constant $g(\Lambda_u,\Lambda_l)$ it is clear that the perturbative correction is not divergent in the ultraviolet (UV) since the limit $\Lambda_u \rightarrow \infty$ can be taken safely and the term $1/\Lambda_u$ goes to zero. This can be understood from Fig.~\ref{perturbative_coupling} where we plot the perturbative coupling $g(\Lambda_u,\Lambda_l)$ normalized by the prefactor $e^2/2m\pi c^2 \epsilon_0 A$ and for a fixed lower cutoff $\Lambda_l$. The coupling constant increases rapidly and asymptotically reaches a fixed value which is $\tfrac{e^2}{2m c^2 \pi \epsilon_0A} \frac{1}{\Lambda_{l}}$ which means that the perturbative multi-mode coupling converges with the UV cutoff $\Lambda_u$. However, the perturbative coupling diverges if the lower cutoff is taken to zero, 
\begin{equation}
\lim_{\Lambda_l \to 0} g(\Lambda_u,\Lambda_l) \rightarrow \infty \, .
\end{equation}
This implies that the perturbative coupling is divergent in the infrared part of the electromagnetic spectrum as we have also seen in the main part of this work. As a consequence the perturbative correction to the free particle dispersion $\Delta E$ becomes arbitrarily negative and thus the perturbative computation leads to an instability as the particle dispersion turns from positive to negative. Thus, perturbation theory violates the boundedness of the Pauli-Fierz Hamiltonian from below and the perturbative free-particle spectrum no longer has a minimum. We note that up to the factor $\pi^2/6$ we obtain the same result as the discretized form of Eq.~\eqref{eq:discretizedperturabtion}. This difference is merely due to the fact that we first perform the infinite sum before performing the limit of $L \rightarrow \infty$ for the integration. For notational simplicity we keep the explicitly discretized form in the main text.  

\begin{figure}
\centering
\includegraphics[width=0.8\linewidth]{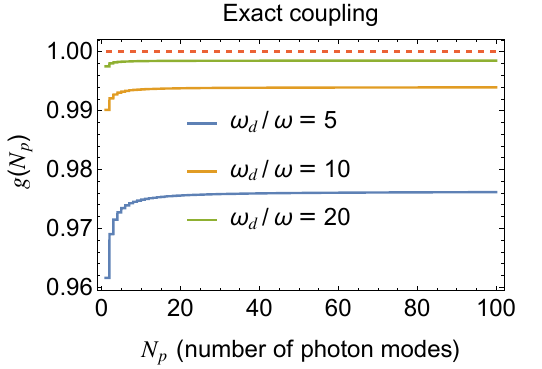}
\caption{Non-perturbative multi-mode coupling $g(N_p)$ as obtained from the exact solution for the free particle in Eq.~(\ref{eq:free-electron-photon-spectrum}) as a function of the number of photon modes $N_p$, for different values of the ratio $\omega_d/\omega$. The coupling constant increases rapidly and asymptotically reaches a plateau. The value of the plateau depends on the ratio $\omega_d/\omega$ but it never goes beyond unity.}
\label{exact_coupling}
\end{figure}

In contrast to the perturbative computation, the physical picture emerging from the non-perturbative solution of the free particle is different~\cite{rokaj2020}. In Fig.~\ref{exact_coupling} we plot the exact non-perturbative multi-mode coupling constant $g(N_p)$ as a function of the number of photon modes $N_p$ as given from Eq.~(\ref{eq:free-electron-photon-spectrum}). We see that $g(N_p)$ has effectively the same dependence with respect to the amount of photon modes as the perturbative coupling $g(\Lambda_u,\Lambda_l)$ with respect to the upper cutoff $\Lambda_u$. They both increase rapidly and then reach a plateau. However, with the respect to the lower photonic cutoff their behaviors are drastically different. The non-perturbative coupling constant $g(N_p)$ has an upper bound and never exceeds 1, even for very large values of the ratio $\omega_d/\omega=5, 10, 20$. We note that $\omega$ here denotes the lowest frequency considered in the photonic spectrum. From Fig.~\ref{exact_coupling} it is clear that if we fix $\omega_d$ then for arbitrarily small $\omega$ the multi-mode coupling can reach unity but never exceeds it. In contrast to the perturbative coupling, the exact coupling never diverges and as consequence the free particle dispersion is stable (positive) and always well defined. This is decisive and fundamental difference between perturbation theory and the exact solution which highlights the importance of non-perturbative treatment of the light-matter interaction.

The fact that the exact coupling does not diverge even for lowest mode going to zero, $\omega \rightarrow 0$, can be understood from the single-mode case ($N_p=1$) where $g(1)$ is given analytically~\cite{rokaj2020},
\begin{equation}
g(1)=\frac{\omega^2_d}{\omega^2+\omega^2_d}.
\end{equation}
The diamagnetic frequency is $\omega^2_d=e^2/(m_e\epsilon_0 AL)$, and the lowest mode for periodic boundary conditions is $\omega=2\pi c/L$. For $L\to \infty$ we have $\omega^2\sim 1/L^2$, which goes faster to zero than $\omega_d^2\sim 1/L$, and consequently $g(1) \to 1$.

Finally, it is important to mention that despite the fact that the free particle dispersion is always well-defined and the coupling bounded, the observable mass $m_e=m/(1-g(N_p))$ as defined in Eq.~(\ref{eq:renormalized-mass}) diverges when $g(N_p)\rightarrow 1$. To tame the diverging $m_e$ in renormalization theory, the bare mass $m$ becomes cutoff-dependent and is promoted into $m(N_{p})$ such that to exactly cancel the diverging term $1/(1-g(N_{p}))$. For that purpose one takes $m(N_{p})=m_e (1-g(N_{p}))$ where $m_e$ is the observbale electron mass. An important feature of our non-perturbative formula for the mass renormalization is that the bare masses is always positive $m(N_{p})\geq 0$ because the total coupling $g(N_{p})\leq 1$ has an upper bound.

\section{Free Electron in the Length Gauge}
\label{sup:HEGlengthgauge}

In this appendix we provide the solution for a single electron coupled to one photon mode in the length gauge. Our purpose is to show that the renormalized dispersion of the electron in velocity and length gauges is the same. The Hamiltonian of one electron interacting with one photon mode in the length gauge is~\cite{rokaj2017} 
\begin{equation}
\hat{H}_{\text{L}} =\frac{\hat{\textbf{p}}^{2}}{2m} +\frac{1}{2}\left[\hat{p}^2+\omega^2\left(\hat{q} \!-\! \frac{e\boldsymbol{\lambda}}{\omega} \cdot \hat{\textbf{r}} \right)^2\right] , \label{eq:length-gauge-hamiltonian-1D}
\end{equation} 
We choose the polarization of the mode to be in the $x$ direction, $\boldsymbol{\lambda}=\lambda\mathbf{e}_x$, and in one spatial dimension we have 
\begin{eqnarray}\label{lengthHz}
	\hat{H}_{\text{L}}=-\frac{\hbar^2}{2m}\frac{\partial^2}{\partial x^2}-\frac{\hbar^2}{2}\frac{\partial^2}{\partial q^2}+\frac{\omega^2}{2}\left( q-\frac{\lambda}{\omega} x\right)^2, 
\end{eqnarray}
First we perform the scaling transformation $\bar x= e\lambda x/\omega$ and we introduce the parameter $\bar m= m \omega^2/\lambda^2$

\begin{eqnarray}\label{lengthHz}
	\hat{H}_L=-\frac{\hbar^2}{2\bar m }\frac{\partial^2}{\partial \bar x^2}-\frac{\hbar^2}{2}\frac{\partial^2}{\partial q^2}+\frac{\omega^2}{2}\left( q-\bar x\right)^2 .
\end{eqnarray}
The Hamiltonian can be solved by going into the mixed coordinates 
\begin{eqnarray}
    w=\frac{\bar m \bar x+ q}{\bar m +1} \;\; \textrm{and}\;\; u=q-\bar x,
\end{eqnarray}
where it takes the simple form
\begin{equation}
    \hat{H}_{\text{L}}=-\frac{\hbar^2}{2(\bar m +1)}\frac{\partial^2}{\partial w^2} -\frac{(\bar m +1)\hbar^2}{2\bar m} \frac{\partial^2}{\partial u^2} +\frac{\omega^2}{2} u^2.
\end{equation}
In the above Hamiltonian we have a freely propagating polaritonic mode along the $w$ coordinate and harmonically confined mode along the $u$ coordinate. The $w$-dependent eigenfunctions are plane waves $f_{k_w}(w)=e^{\textrm{i}k_ww}$ while the eigenfunctions of the $u$ mode are Hermite functions $\Phi_n(u)$. Then, the energy spectrum of the system is
\begin{equation}\label{length spectrum}
    E_{k_w,n}=\frac{\hbar^2k^2_w}{\bar m +1} +\hbar \widetilde{\omega}\left(n+\frac{1}{2}\right).
\end{equation}
We note that the coordinates $w$ and $u$ are independent as they mutually commute $[\partial_w,\partial_u]=[\partial_w,u]=[\partial_u,w]=0$. Comparing now the spectrum above of the free electron in the length gauge to the one derived in the velocity gauge given in Eq.~(\ref{eq:free-electron-photon-spectrum}) we see that they are not exactly the same. The length gauge spectrum $E_{k_w,n}$ depends on the polaritonic quantum number $k_w$ while the spectrum in the velocity gauge on the quantum number $k_x$. Naturally, the question that arises is: How are $k_x$ and $k_w$ related?

To figure this out we will use the relation between the differential operators of $\partial_x$ and $\partial_w$. From the chain rule and neglecting the contribution of the photonic coordinate $q$ we have
\begin{eqnarray}\label{kw and kx}
\frac{\partial}{\partial w}=\frac{\partial \bar x}{\partial w} \frac{\partial x}{\partial \bar x}\ \frac{\partial}{\partial  x}=\frac{\omega}{\lambda}\frac{\partial}{\partial x}\; \Longrightarrow\; k_w=\frac{\omega}{\lambda}k_x.
\end{eqnarray}
Substituting the relation above into Eq.~(\ref{length spectrum}) we find for the length gauge spectrum
\begin{eqnarray}
E_{k_x,n}=\frac{\hbar^2 k^2_x}{2m}\left(1-\frac{\omega^2_d}{\widetilde{\omega}^2}\right)+\hbar\widetilde{\omega}\left(n+\frac{1}{2}\right).
\end{eqnarray}
The above result reproduces precisely the single-particle dispersion coupled to a single photon mode obtained in the velocity gauge in Ref.~\cite{rokaj2020}. This shows that the same free particle dispersion and the corresponding renormalized mass can be consistently obtained from both gauges.

\begin{figure}[bth]
\includegraphics[width=1.0\columnwidth]{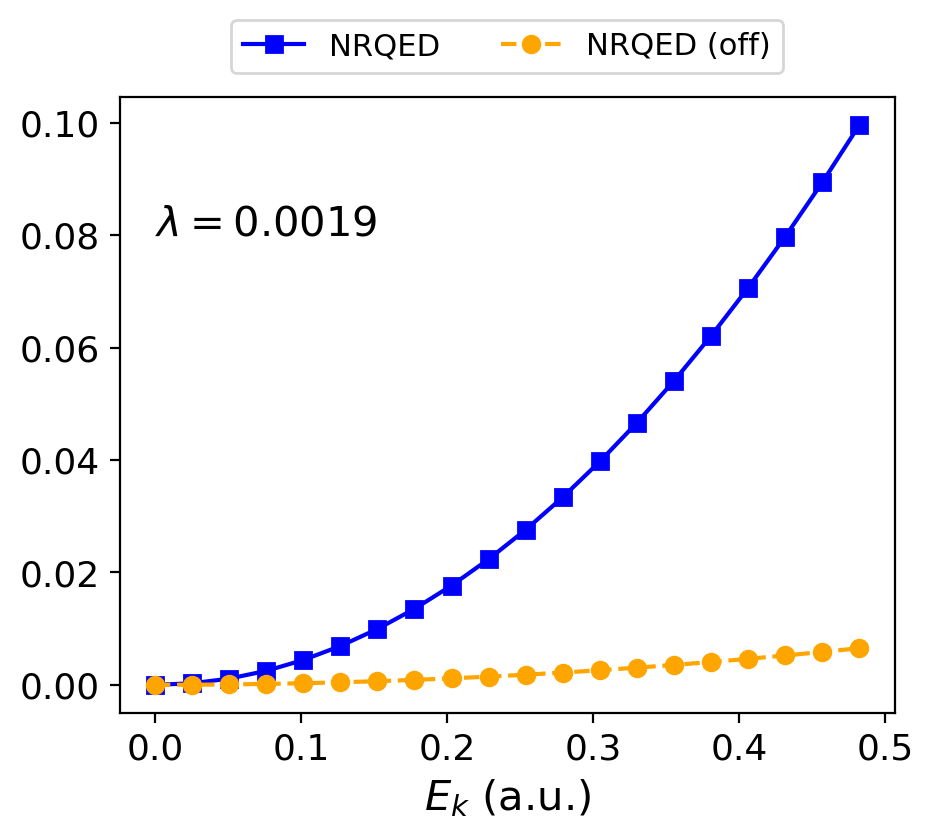}
\caption{A comparison of the energy dispersion between two different sampling of the discretized photonic continuum. NRQED is the result presented in Fig.~(\ref{fig:free-electron-M-modes}a) and NRQED (off) is the case where the upper cutoff is $0.05$~au.}
\label{fig:dispersion-energy-compare}
\end{figure}

We now revisit the issue of a mismatch between light and matter if both systems are not chosen consistently. To illustrate this, we keep a fixed length scale for the free particle as done in Sec.~\ref{sec:free-electron} and also keep the same sampling of 200 photon modes with cutoffs 0.01 and 0.5 au. We now sample a different discretized continuum with 200 photon modes but with cutoffs 0.001 and 0.05 au. Here, the upper cutoff is much lower than the energy of the first excitation of the free particle. For both continua, the coupling of the photon modes to the free particle is fixed to $\lambda=0.0019$. A comparison of the energy dispersion is shown in Fig.~(\ref{fig:dispersion-energy-compare}) where we find that the NRQED case with the upper cutoff (0.05 au) is off from the NRQED case with upper cutoff (0.5 au). The reason for this mismatch is that the photonic modes are all excited before the first electronic state can be populated making the matter degrees less important in the coupled system. This result shows that choosing length-scales consistently is very important in QED.

\section{Model of a one-dimensional {atom} }
\label{sup:1D-atomic-hydrogen}

The one-dimensional atomic system we consider features a single bound electron. The quantum mechanical Hamiltonian describing this system is given by
\begin{align}
\hat{H}_{\text{atom}} &= - \frac{\hbar^{2}}{2m_{e}}\frac{\partial^{2} }{\partial x^{2}} -\frac{e^{2}}{4\pi\epsilon_{0}}\frac{v_{0}}{\cosh^{2}(k_{0}x)} \,  . \label{eq:qm-hamiltonian}
\end{align}
where $v_{0}$ and $k_{0}$ are parameters that control the depth of the P\"{o}schl-Teller potential~\cite{Poschl1933}. For a single electron in the binding potential, the analytic spectrum of Eq.~(\ref{eq:qm-hamiltonian}) is given as~\cite{landau1977}
\begin{align}
E_{n} &= - \frac{\hbar^{2}k_{0}^{2}}{8m_{e}} \left(\sqrt{1 + \frac{8m_{e}v_{0}k_{e}}{\hbar^{2}k_{0}^{2}}} - (1 + 2n) \right)^{2} , \label{eq:analytic-spectrum}
\end{align}
where the quantum numbers are $n \in \mathbb{N}$ and $k_{e}=e^{2}/4\pi\epsilon_{0}$. The number of bound states can be controlled using $v_{0}$ and $k_{0}$. For our calculations we choose $v_{0}=10$ and $k_{0}=0.05$ which gives 10 bound states of interest. We use the analytic results to benchmark our numerical implementation which quantitatively agree.

\section{Model of a one-dimensional H$_{2}$ molecule}
\label{sup:1D-molecular-hydrogen}



\begin{figure}[bth]
\includegraphics[width=1.0\columnwidth]{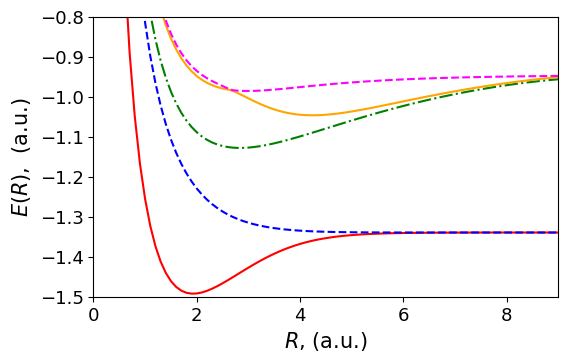}
\caption{The first five numerically exact PESs of the 1D hydrogen molecule H$_{2}$ for $N_{p}=0$. The harmonic approximation to the ground-state PES has the harmonic frequency $\omega_{e}=0.020455$ a.u. for the nuclear degrees.}
\label{fig:PES-1D-H2-molecule}
\end{figure}

The Hamiltonian of the model system of the H$_{2}$ molecule in one-dimension where the relevant coordinates are the internuclear separation, $R$, and the electronic coordinates, $x_{1}$ and $x_{2}$ is given by
\begin{align}
\hat{H}_{\text{mol}} &= - \frac{1}{2\mu_{n}}\frac{\partial^{2} }{\partial R^{2}} + \frac{1}{R} + \frac{1}{\sqrt{(x_{1}-x_{2})^{2} + a_{\text{ee}}}} \nonumber \\
& \quad + \sum_{i=1}^{2} \left(- \frac{1}{2\mu_{\text{e}}}\frac{\partial^{2} }{\partial x_{i}^{2}} -\frac{1}{\sqrt{(x_{i}-R/2)^{2} + a_{\text{en}}}} \right .\\
&\qquad\qquad\quad \left .  - \frac{1}{\sqrt{(x_{i}+R/2)^{2} + a_{\text{en}}}}\right) , \nonumber
\label{eq:1D-H2-hamiltonian}
\end{align}
where $\mu_{\text{e}}=2M_{n}/(2M_{n}+1)$ and $\mu_{n}=M_{n}/2$ are the reduced observable electronic and nuclei masses, respectively. We take the proton mass to be $M_{n}= 1836 \, m_{e}$. The electron-electron and electron-nuclear interaction terms are represented by soft-Coulomb potentials where the soft-Coulomb parameters take values $a_{\text{ee}}\!=\!2$ and $a_{\text{en}}\!=\!1$. For the model, the PESs are defined by the following electronic eigenvalue problem: $\hat{H}_{\text{el}}(x_{1},x_{2};R)\Phi_{k}(x_{1},x_{2};R)=E_{k}(R)\Phi_{k}(x_{1},x_{2};R)$ where $\hat{H}_{\text{el}}=\hat{H}_{\text{mol}} - \hat{T}_{n}$ where $\hat{T}_{n}=-\frac{1}{2\mu_{n}}\frac{\partial^{2} }{\partial R^{2}}$. We show the first five numerically exact PESs in Fig.~(\ref{fig:PES-1D-H2-molecule}) for the case where we do not couple to the photonic continuum (i.e. for $N_{p}=0$). The mean nuclear equilibrium position is $R_{\text{eq}}=1.9$ a.u. with the corresponding ground-state energy $E_{0}=-1.4843$ a.u. Applying the harmonic approximation to the ground-state PES as in App.~\ref{sup:morse-fit-PES} we obtain the harmonic frequency $\omega_{e}=0.020455$ a.u. of the nuclear degrees.

\begin{figure}[bth]
\includegraphics[width=1.0\columnwidth]{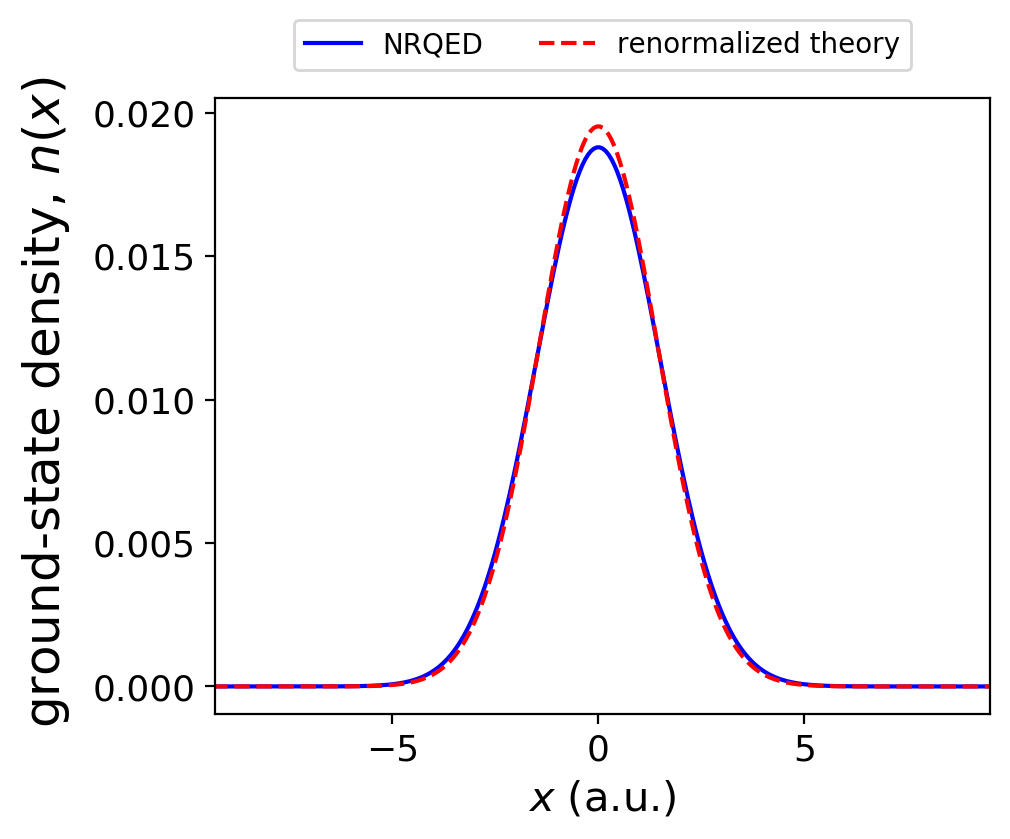}
\caption{A comparison of the ground-state density of the H$_{2}$ molecule at the equilibrium position $R_{\text{eq}}=1.9$ a.u. between NRQED and {the free-space renormalized theory} for the case of $N_{p}=15$ photon modes.}
\label{fig:gs-density-QM-vs-bQM}
\end{figure}

We note that when we couple the molecule to the discretized continuum, we used the bare proton mass $M=1836 \, m$ where the bare electronic mass is $m=1$. At the equilibrium position, we compute the ground-state density for NRQED and {the renormalized theory} for the case when the renormalized mass is obtained with the lowest 50 of the 200 sampled photon modes. In Fig.~(\ref{fig:gs-density-QM-vs-bQM}), we show the ground-state density where the {free-space renormalized} case is more bound when compared to the NRQED. The reason for this is discussed in Sec.~\ref{subsec:atomic-system} of the main text.

\section{Morse and harmonic approximation to the H$_{2}$ PES}
\label{sup:morse-fit-PES}

\begin{figure}[bth]
\includegraphics[width=1.0\columnwidth]{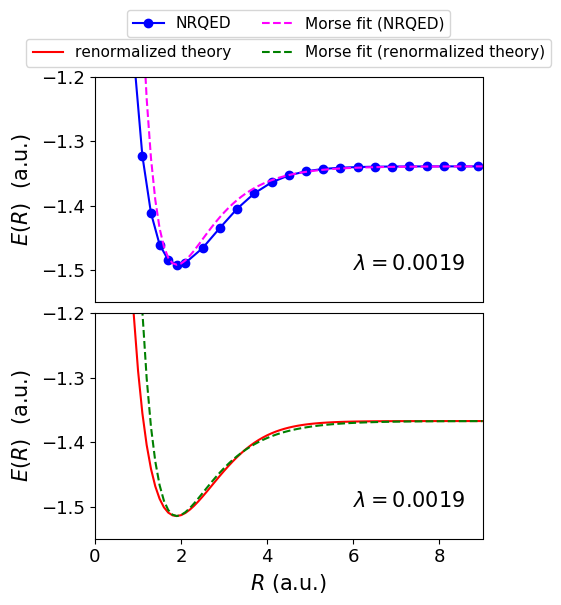}
\caption{Morse potential fit to the numerical exact ground-state PES of the H$_{2}$ molecule for NRQED and {the free-space renormalized theory} for the case of $N_{p}=15$ photon modes.}
\label{fig:BOPES-NRQED-morse-fit}
\end{figure}

In this section we provide details of the Morse and harmonic approximation to the numerical exact ground-state H$_{2}$ PoPES of the NRQED and {the renormalized} settings. To do this, we first consider the Morse potential
\begin{equation} 
V_{\text{M},i}(R) = D_{e,i}\left( e^{-a_{i}(R - R_{\text{eq}})} - 1 \right)^{2} + C_{i} \, , 
\label{eq:morse-potential}
\end{equation}
where the parameter ``$a$" controls the `width' of the potential (i.e., the smaller ``$a$" is, the larger the well), $C$ is a constant shift in the PES and $i = 
\{\textrm{NRQED}, \textrm{renormalized}\} $. Since we have access to all the parameters of Eq.~(\ref{eq:morse-potential}) except for the $a$ parameter, this makes the fitting procedure easier. To fit the Morse potential to the exact results of NRQED and {the free-space renormalized} settings, we employ the ``curve\_fit'' function of scipy and the corresponding $a$ parameter values are given in Tab.~(\ref{tab:morse-a-parameter}). The results of the fit are plotted in Fig.~(\ref{fig:BOPES-NRQED-morse-fit}).
\begin{table}[bth]
\begin{tabular}{ | c | c | c |  }
    \hline
    Level of theory  & Morse parameter $a$  & Harmonic frequency $\omega_{e}$ \\
    \hline 
    {renormalized}       &  $1.12908125$  &  $0.0202418$ \\
    \hline
    NRQED    &  $1.18930095$  &  $0.0217007$  \\
    \hline
\end{tabular}
\caption{The values of the $a$ parameter of the Morse potential resulting from the fit and the deduced harmonic frequency $\omega_{e}$ for the case of $N_{p}=15$ photon modes. The units are in (a.u.).}
\label{tab:morse-a-parameter}
\end{table}
Since we are interested in the influence the continuum has on the nuclear degrees, we connect the $a$ parameter to the nuclei mass by employing the harmonic potential fit to the Morse potential around the equilibrium $R_{\text{eq}}$. The harmonic potential is given by
\begin{equation} 
V_{\text{H},i}(R) = \frac{1}{2}k_{i}\left( R - R_{\text{eq}} \right)^{2} + C_{i} \, , 
\label{eq:harmonic-potential}
\end{equation}
where $k$ is the force constant of the bond which is related to the reduced nuclei mass as $k = \mu_{n}\, \omega_{e}^{2}$ and $\omega_{e}$ is the vibrational frequency of the potential. From the above considerations, we have the relation $k=2D_{e} a^{2}$ from which we have $\omega_{e} = \sqrt{2D_{e} a^{2}/\mu_{n}}$.

To obtain the approximate vibrational harmonic frequency of the NRQED setting, we used the bare proton mass $M=1836 \, m$ where the bare electronic mass is $m=1$. The renormalized proton mass is deduced from the energy dispersion for free interacting protons coupled to the electromagnetic continuum. The energy dispersion is similar to Eq.~(\ref{eq:free-electron-photon-spectrum}) where the diamagnetic frequency has a dependence on the nuclear charge.

\vspace{10em}

\bibliography{01_light_matter_coupling}

\end{document}